# Beyond Yield: Integrating Energy, Water, Cost, and Carbon to Benchmark Indoor Vertical Farming Viability

Francesco Ceccanti[1]*, Aldo Bischi[1], Umberto Desideri[1], Andrea Baccioli[1]

1 Department of Energy, System, Territory and Construction Engineering
Università di Pisa, Pisa, Italy

*Corresponding Author: francesco.ceccanti@phd.unipi.it (F. Ceccanti)

**Abstract**

The interest in vertical farming arises from its ability to ensure consistent, high-quality, and pest-free vegetable production while supporting synergies with energy systems and urban development. While previous studies have assessed energy use and cost-effectiveness in vertical farming using simplified models, this study fills a gap by providing a comprehensive analysis of how individual input parameters affect efficiency, sustainability, and economic viability through a detailed modeling framework with sensitivity and correlation analyses. 162 scenarios were evaluated by combining three levels of temperature, photosynthetic photon flux density, and $CO_2$ concentration across three distinct localities, namely Trondheim, Shanghai, and Dubai regions, which differ from a socio-environmental viewpoint. Two insulation thicknesses were also tested in each scenario. Results indicate that, due to the heating, ventilation, and air conditioning and dehumidification system, crop productivity could be kept optimal regardless of insulation or the external climate. Photosynthetic photon flux density was the dominant growth factor (correlation: 0.85), followed by $CO_2$ (0.36) and temperature (0.22), and also the driver of energy consumption (0.73). The lowest specific energy consumption coincided with the lowest productivity (55 $kg\,m^{-2}$). Levelized cost of lettuce identified the most cost-effective setup as 24 °C, 250 $\mu mol\,m^{-2}\,s^{-1}$ photosynthetic photon flux density, 1400 ppm $CO_2$, with insulation (102 $kg\,m^{-2}$. Only decarbonized energy systems can support vertical farming without increasing $CO_2$ emissions compared to imported lettuce. These findings guide practitioners and policymakers in selecting cost-effective, sustainable vertical farming strategies and provide validated data for research and implementation across diverse climatic and energy contexts.

Nomenclature

| **Acronyms** | | | **Subscripts** | |
|---|---|---|---|---|
| **AHU** | Air Handling Unit | [–] | **ahu** | Air Handling Unit |
| **CAPEX** | Capital Expenditures | [$] | **air** | Internal Air |
| **CEA** | Controlled Environment Agriculture | [–] | **c** | Crop |

| **COP** | Coefficient of Performance | $[kW_{th}\, kW_{el}^{-1}]$ | **CO₂** | Carbon Dioxide |
|---|---|---|---|---|
| **CRF** | Capital Recovery Factor | $[-]$ | **cond** | Condensed/Condensation |
| **DLI** | Daily Light Integral | $[\mu mol\, m^{-2}\, day^{-1}]$ | **dm** | Dry Matter |
| **DMC** | Dry Matter Content | $[-]$ | **el** | electrical |
| **HVAC** | Heating, Ventilation and Air Conditioning | $[-]$ | **ET** | Evapotranspiration |
| **HVACD** | Heating, Ventilation and Air Conditioning and Dehumidification | $[-]$ | **env** | Envelope |
| **LAI** | Leaf Area Index | $[{m_{leaf}}^2\, {m_{soil}}^{-2}]$ | **eva** | Evaporation |
| **LCoL** | Levelized Cost of Lettuce | $[\$\, kg^{-1}]$ | **ext** | External |
| **LUE** | Light Use Efficiency | $[g\, \mu mol^{-1}]$ | **fm** | Fresh Matter |
| **OPEX** | Operational Expenditure | $[\$\, year^{-1}]$ | **H/C** | Heating & Cooling Systems |
| **PAR** | Photosynthetic Active Radiation | $[\mu mol\, m^{-2}\, s^{-1}]$ | **hum** | Humidification System |
| **PLF** | Part-Load Factor | $[-]$ | **HP** | Heat Pump |
| **PLR** | Part-Load Ratio | $[-]$ | **in** | Internal |
| **PPFD** | Photosynthetic Photon Flux Density | $[\mu mol\, m^{-2}\, s^{-1}]$ | **ins** | Insulation |
| **SEC** | Specific Energy Consumption | $[kWh\, kg^{-1}]$ | **Lab** | Labor |
| **VF** | Vertical Farm | $[-]$ | **Leas** | Leasing |
| **VPD** | Vapour Pressure Deficit | $[kPa]$ | **LED** | Lighting System |
| **WUE** | Water Use Efficiency | $[g_{FM}\, L^{-1}]$ | **n** | Net |
| **Symbols** | | | **Nom** | Nominal |
| $A$ | Area | $[m^2]$ | **plant** | Single Lettuce Plant |
| $c$ | Specific Cost | $[\$\, x^{-1}]$ | **rad** | Radiative |
| $c_p$ | Specific Heat Capacity | $[J\, kg^{-1}\, K^{-1}]$ | **Rep** | Replacement |
| $C$ | Cost | $[\$]$ | **T** | Temperature-Dependent |
| $d$ | Density | $[kg\, m^{-3}]$ | **tot** | Annual |
| $D$ | Rank Difference | $[-]$ | **wat** | Water |
| $DM$ | Dry Matter | $[g\, m^{-2}]$ | **wir** | Wiring |
| $e$ | Specific Emissions | $[g\, ton^{-1}\, km^{-1}]$ | | |
| $F$ | Photosynthetic Photon Efficiency | $[\mu mol\, J^{-1}]$ | | |
| $f$ | $CO_2$ factor for crop growth | $[-]$ | | |
| $FM$ | Fresh Matter | $[g\, m^{-2}]$ | | |
| $K$ | Evapotranspiration Coefficient | $[-]$ | | |
| $i$ | Inflation Rate | $[-]$ | | |
| $I$ | Solar Irradiation | $[W\, m^{-2}]$ | | |
| $k$ | Light Extinction Coefficient | $[-]$ | | |
| $L$ | Supply Distance | $[km]$ | | |
| $M$ | Mass | $[kg]$ | | |
| $MW$ | Molecular Weight of Glucose | $[g\, mol^{-1}]$ | | |
| $\dot{m}$ | Mass Flow | $[kg\, s^{-1}]$ | | |
| $n$ | Number of Observations | $[-]$ | | |
| $P$ | Electrical Power | $[kW]$ | | |
| $Q$ | Thermal Power | $[kW]$ | | |
| $r$ | Interest Rate | $[-]$ | | |
| $RAD$ | Short-wave Radiation | $[MJ\, m^{-2}\, day^{-1}]$ | | |
| $RF$ | Root Factor | $[-]$ | | |

| $\boldsymbol{T}$ | Temperature | $[°K]$ | | |
|---|---|---|---|---|
| $\boldsymbol{t}$ | Time | $[s]$ | | |
| $\boldsymbol{u}$ | Air Speed | $[m\ s^{-1}]$ | | |
| $\boldsymbol{U}$ | Thermal Transmittance | $[W\ m^{-2}\ K^{-1}]$ | | |
| $\boldsymbol{V}$ | Volume | $[m^3]$ | | |
| $\boldsymbol{\alpha}$ | Wall's Absorption Factor | $[-]$ | | |
| $\boldsymbol{\varepsilon}$ | Wall's Emissivity Factor | $[-]$ | | |
| $\boldsymbol{\gamma}$ | Psychrometric constant | $[kPa\ °C^{-1}]$ | | |
| $\boldsymbol{\rho}$ | Spearman Coefficient | $[-]$ | | |
| $\boldsymbol{\eta}$ | Efficiency | $[-]$ | | |
| $\boldsymbol{\Delta}$ | Slope of the Saturation Curve | $[kPa\ °C^{-1}]$ | | |
| $\boldsymbol{\Delta E_{phot}}$ | Specific Energy Requirement of Photosynthesis Reaction | $[J\ mol^{-1}]$ | | |

## 1. Introduction

In recent years, the increasing need for sustainable, high-quality food production has highlighted controlled-environment agriculture (CEA) as a modern and highly relevant solution. Vertical farms (VFs) are food production systems that use hydroponic or aeroponic methods, with crops grown in multiple layers within enclosed spaces like buildings, containers, or cells. In these systems, key growth factors such as temperature, humidity, carbon dioxide levels, and photosynthetically active radiation (PAR) are meticulously controlled to optimise crop growth [1].

### 1.1. Vertical Farming: Concepts and Benefits

Vertical farming increases yields dramatically by exploiting vertical space, achieving ten to one hundred times higher production than open field farming [2] and proportional gians compared to greenhouses [3] on the same land area. Controlled conditions not only accelerate growth but also improve light use efficiency (LUE): average LUE values of 0.55 $g_{dw}\ mol^{-1}$ have been reported, compared to 0.39 $g_{dw}\ mol^{-1}$ in greenhouses and 0.23 $g_{dw}\ mol^{-1}$ in open fields, with peaks up to 1.63 $g_{dw}\ mol^{-1}$, close to the theoretical limit of 1.81 $g_{dw}\ mol^{-1}$ [4]. Experiments on lettuce [5] 0.8 to 1.25 $g_{dw}\ mol^{-1}$), highlighting the strong influence of photoperiod, daily light integral (DLI), spectral composition, $CO_2$ concentration, temperature, and photosynthetic photn flux density (PPFD) [6].

Hydroponics is currently the most widespread VF method, thanks to its relatively low costs, ease of operation, reduced risk of soil borne diseases [7], no pesticide requirements, and high resource efficiency, with water savings of seventy to ninety five percent compared to conventional farming [2]. These advantages supported the sector's rapid growth from USD 2.14 billion in 2018 to USD 6.20 billion in 2023 [7], motivating the present study's focus on hydroponic vertical farming.

### 1.2. Vertical Farming: Challenges and Limitations

Vertical farming faces two major limitations: its high energy requirements and the restricted range of crop types that can be effectively cultivated in such systems. The energy consumption of vertical farms (VFs) has been widely investigated, with reported values ranging from 3.2 to 20 kWh $kg^{-1}$ depending on the specifications of LED lighting, HVACD systems, and indoor conditions [8]. This significant variability highlights the need for a more detailed investigation into how specific input parameters influence the overall efficiency and performance of VFs. A recent review compared the energy demand of VFs with that of closed and open greenhouses, showing that VFs consume approximately twice as much energy as closed greenhouses and about three times more than open greenhouses to produce the same amount of lettuce. Notably, around 70% of a VF's total energy use is attributed to the lighting system, while heating, ventilation, and air conditioning (HVACD) systems account for approximately 28% [2]. Given that lighting accounts for most of the energy consumption in vertical farms, [9] provides a comparative analysis of various vegetable crops based on their photon cost per unit of dry mass. The study also evaluates the photon cost as a percentage of the estimated market price of the corresponding dry mass, to identify the most cost-effective plant species for vertical farming. Crops characterized by a high harvest index (HI), high water content, and relatively high market value, such as lettuce, leafy microgreens, and tomatoes, are shown to be economically viable under artificial lighting. In particular, the photon cost associated with lettuce cultivation amounted to 5% of its corresponding dry market price, which helps explain why lettuce is the most commonly selected crop in VF-related studies. In contrast, [9] demonstrates that even with 100% efficient LED systems, the cultivation of staple crops like rice and wheat in vertical farms remains economically unfeasible. The widespread popularity of lettuce led the authors to select it as the reference crop for the VF system analyzed in detail in the present study.

### 1.3. Recent Studies

Recent studies aiming to improve the specific energy consumption (SEC) of vertical farms have primarily focused on two main aspects: reducing the overall energy demand of the system, and optimizing internal growing conditions to enhance crop productivity and, consequently, energy efficiency.

To decrease the overall energy demand, various strategies have been explored. A systematic review [2] categorizes Energy Efficiency Measures (EEMs) into three main groups: building architecture and envelope, distributed generation, and HVACD systems. Among these, photovoltaic (PV) arrays have been identified as a cost-effective energy source for CEA applications. Moreover, passive heating and cooling strategies, leveraging the local climate conditions, can reduce HVACD energy loads by up to 31%.

Integration with energy systems and urban infrastructure further enhances the competitiveness of vertical farms. For instance, combining vertical farming with agrivoltaics systems (VFCA) has demonstrated up to a 13-fold yield increase compared to conventional closed agrivoltaics [10]. Studies have also assessed fully PV-powered vertical farms, showing that the required PV surface areas vary depending on the location [11]. From an urban perspective, integrating vertical farms into architectural designs can increase local food production while providing social benefits [12]. Façade-based systems have been evaluated for both energy performance and economic feasibility relative to conventional farming [13].

In parallel, BIPV-driven "smart" vertical greenhouses [14] and agrivoltaics building envelopes [15] are emerging to co-produce food and electricity in cities and to embed vertical farming within Water-Energy-Food-Environment (WEFE) nexus strategies, revealing notable synergies and design trade-offs at façade or greenhouse scale. In this context, the current study's contribution provides a validated, dynamic coupling from agronomic set-points to energy–cost–carbon outcomes in fully controlled indoor VFs with explicit envelope/HVACD modelling and multi-climate generalization.

### 1.4. Research Gap

As highlighted by the existing literature, interest in vertical farming is rapidly increasing, partly due to its multidisciplinary potential across agriculture, energy systems, and urban integration. Hence, developing best practice guidelines for the design and management of vertical farms, together with a deeper understanding of how key parameters influence VF performance, is becoming essential to reduce both energy consumption and operational costs. Achieving this goal requires a more comprehensive modelling framework and methodological approach.

In 2021, Weidner et al. [16] compared vertical farming with open and closed greenhouses by optimizing the indoor conditions to minimize the specific energy consumption. However, to enable the optimization over a 24-hour period, the crop component of the model, including growth and evapotranspiration, was oversimplified, assuming constant values for some key growth parameters. On the economic side, Moghimi et al. [17] developed a methodology to compare the total production costs of vertical farming with those of conventional agriculture across seven U.S. states, providing a regionalized economic assessment. While both studies offered a detailed comparison of energy use and economics relative to traditional farming, their assumptions on resource consumption and productivity in vertical farming lack a rigorous and comprehensive modelling framework, limiting a deeper understanding of VF performance. Indeed, most of the studies, including the energy consumption of vertical farming, tend to assume constant values to estimate both crop growth and evapotranspiration. More recently, in 2024, [18] introduced an integrative study combining both agronomic and economic perspectives. This work assessed the effects of indoor temperature, DLI, and LED efficiency on crop yield, energy consumption, and the cost of lettuce production, with the analysis extended across three distinct climatic zones. The findings underline the importance of site-specific design and operation strategies to ensure the sustainability and profitability of vertical farming systems. However, in [18], neither the impact of the $CO_2$ concentration on crop growth nor the effect of the envelope's insulation on energy consumption was investigated. Furthermore, that study's cost analysis for lettuce considered only electricity and natural gas expenses. Ultimately, [19] conducted a carbon emissions analysis related to vertical lettuce cultivation, based on a limited case study conducted in the Netherlands. Hence, although several studies have assessed the energy use and cost-effectiveness of vertical farming systems, the literature still lacks a comprehensive analysis that provides guidelines and insights into how each of the input parameters affects VF performance within a detailed and rigorous modelling framework.

### 1.5. Study's Contribution

The scientific contribution of this study lies in offering a comprehensive analysis of production performance and energy consumption in vertical farming systems, intending to clarify how individual input parameters affect efficiency (expressed in kWh $kg^{-1}$), sustainability (expressed in $kg_{CO2}$ saved per year), and economic viability (expressed in \$ $kg^{-1}$), using a sensitivity and a distance correlation analysis. This approach is more appropriate than the Spearman correlation coefficient for capturing complex nonlinear relationships. The results of the analysis were then used to develop some best practices for the design and management of a VF, tailored to different socio-environmental conditions.

This study builds on [20] by integrating temperature, photosynthetically active radiation, and $CO_2$ concentration into a validated dynamic energy model to assess their impact on energy use [21]. Unlike the transient model in [22], limited to thermal load and crop yield, the adopted framework evaluates thermal load, crop production, water, and $CO_2$ requirements, and electricity consumption. Indoor conditions are assumed to be maintained by a heat-pump-based HVACD system, consistent with CEA electrification goals [23]. The model also incorporates heat transfer through the building envelope, enabling the analysis of VF thermal behaviour under different climates, an extension beyond [22], where indoor temperature was fixed. Economically, this work considers both capital expenditures CAPEX) and a detailed operational expenditures (OPEX) assessment to estimate lettuce production costs, improving on [18]. Finally, the study evaluates the carbon footprint, benchmarking against a non-local lettuce supply chain across multiple climate zones to provide critical performance indicators.

This work simulates 162 scenarios spanning three levels of temperature, PPFD, and $CO_2$ across three climatic/social zones, with two insulation options; quantifies electricity use under a heat-pump-based HVACD consistent with CEA electrification; evaluates carbon footprint, and levelized production cost for lettuce under location-specific mixes; and applies distance correlation to capture non-linear dependencies, distilling best-practice guidelines for VF design and operation.

## 2. Methods

The methods employed for the study of the aforementioned scenarios are detailed in this section. It is divided into three subsections covering the Agri-Energy model for simulating the vertical farm, the studied cases, and the economic assumptions.

### 2.1. Vertical Farm Agri-Energy Model

The assessments extensively discussed in the following Results section are based on the comprehensive agri-energy model presented by Ceccanti et al. [21]. The model depicted in Figure 1 was specifically developed to quantify the energy requirements associated with vertical farm operations. The model incorporates the simulation of all major subsystems involved in vertical farm operation, including artificial lighting, humidification, and dehumidification, HVACD, and $CO_2$ enrichment. Environmental control is achieved by modulating key variables to evaluate their impact on both energy demand and crop productivity. Indoor air temperature is maintained at the defined setpoint via heating and cooling systems, while the desired photosynthetic photon flux density (PPFD) is ensured through LED-based lighting. A spray-based humidification system and an Air Handling Unit (AHU) regulate the relative humidity (RH), keeping it within the optimal range for plant growth. $CO_2$ concentration is controlled through an enrichment system comprising gas cylinders and modulating valves, designed to offset $CO_2$ uptake by plants and losses through ventilation. To mitigate the accumulation of phytotoxic compounds and ensure air quality, a minimum air exchange rate is provided by the ventilation and filtration system. This comprehensive configuration allows for an accurate assessment of the interactions between environmental control strategies and overall vertical farm performance.

The energy balance within the growing chamber is described in Eq. (1), where each Q-term represents a specific energy flow associated with the subsystems described above. Specifically, $Q_{Light}$ denotes the radiant energy supplied by the LED lighting system, while $Q_{plant}$ corresponds to the fraction of this energy absorbed and effectively converted by the crop through photosynthesis. The thermal loads $Q_{h/c}$ and $Q_{AHU}$ are energy loads associated with the Heating/Cooling (HC) system and the AHU, respectively. $Q_{eva}$ and $Q_{hum}$ represent latent cooling loads associated with the evaporation of water transpired by the crop and introduced by the humidification system, respectively. Lastly, $Q_{env}$ accounts for heat exchange with the external environment through the chamber envelope, encompassing both thermal losses and gains.

$$d_{air} c_{p_{air}} V_{room} \frac{\partial T}{\partial t} = Q_{env} + Q_{Light} - Q_{plant} - Q_{eva} + Q_{h/c} - Q_{AHU} - Q_{hum} \quad (1)$$

All these energy flows are extensively detailed in the reference modelling paper [21]. Therefore, only the key equations (2), (3), (4), and (4) are reported in this section to provide the essential energy-related insights. In particular, $Q_{rad}$ accounts for the net heat transfer through the building envelope due to both incident solar radiation and longwave infrared radiation exchange with the sky. $Q_{Light}$ includes the total energy emitted by the lighting system, comprising both the luminous component and the associated thermal load. A portion of the luminous energy is absorbed by the crop canopy ($Q_{abs}$), while the remainder is either reflected or converted into thermal energy via evapotranspiration (ET), thereby contributing to the internal thermal load. The fraction of light effectively used for biomass production was estimated by assuming that the change in dry matter (DM) corresponds to glucose synthesis via the photosynthetic reaction.

$$Q_{env} = U_{wall}A_{wall}(T_{in} - T_{ext}) + Q_{rad} \quad (2)$$

$$Q_{rad} = \alpha I + \sigma\varepsilon A(T_{wall}^4 - T_{sky}^4) \quad (3)$$

$$Q_{Light} = \frac{\frac{PAR}{F} \cdot A_c}{\eta_{LED}} \quad (4)$$

$$Q_{plant} = \frac{\frac{\partial DM}{\partial t}}{MW} A_c \cdot \Delta E_{photo} \quad (5)$$

The dynamic growth model adopted for lettuce cultivation, previously calibrated and validated [21], is primarily governed by equations (6) and (7), which describe the instantaneous variation in dry matter (DM) and fresh matter (FM) as functions of key environmental growth factors, namely temperature, $CO_2$ concentration, and PPFD. At each simulation timestep, the leaf area index (LAI) was computed based on the DM from the previous timestep, a root-to-shoot allocation factor, and the specific leaf area (SLA), which is itself influenced by the prevailing indoor environmental conditions. As the original model was calibrated under a fixed $CO_2$ concentration of 1200 ppm, an additional correction factor was introduced in this work to account for variations in $CO_2$ levels. This factor, derived from [24], accounts for the impact of $CO_2$ concentration on light use efficiency (LUE) as a function of the selected PPFD, thereby enabling a more accurate representation of crop productivity across different atmospheric enrichment scenarios.

$$\frac{dDM}{dt} = PAR \cdot (1 - e^{-k \cdot LAI}) \cdot LUE_{dm} \cdot A_c \cdot f_{CO_2} \quad (6)$$

$$\frac{dFM}{dt} = PAR \cdot (1 - e^{-k \cdot LAI}) \cdot LUE_{fm}(t) \cdot A_c \cdot f_{CO_2} \quad (7)$$

To evaluate the water demand of the VF system, Eq. (8) was applied to estimate the water vapor flux transpired from the leaf surface of lettuce plants. The crop coefficient ($K_c$) was experimentally determined in [25] and varies according to the crop's phenological stage. Additional methodological details on this approach are comprehensively discussed in the referenced model study [21]. The total water demand was then quantified by summing the moisture removed by the cooling and AHU systems ($m_{cond}$), the water content retained in the harvested fresh biomass, and the amount of water sprayed by the humidification system ($m_{hum}$), which is especially relevant during the early growth stages to maintain optimal relative humidity, as shown in Equation (9). Finally, the $CO_2$ uptake by the crop was estimated based on a stoichiometric assumption, whereby the amount of carbon fixed is equivalent to twice the mass of dry biomass produced, following the methodology outlined in [26].

$$\dot{m}_{ET} = 1.05 \cdot \frac{0.408 \cdot \Delta \cdot RAD_n + \frac{900 \cdot u \cdot \gamma \cdot VPD_{in}}{T_{in} + 273.15}}{\Delta + \gamma \cdot (1 + 0.34 \cdot u)} \cdot A_c \cdot K_c \quad (8)$$

$$M_{wat} = \int_{year} (\dot{m}_{ET} + \dot{m}_{hum} - \dot{m}_{cond})dt + (FM_{tot} - DM_{tot}) \quad (9)$$

All the values for the key parameters presented in these equations are summarized in Table *1*.

Table 1: *Main assumptions for the VF transient model presented in [21]*

| Symbol | Value | Unit |
|---|---|---|
| $U_{wall}$ | 0.193 | $W\ m^{-2}\ K^{-1}$ |
| $F$ | 3 | $\mu mol \cdot J^{-1}$ |
| $\eta_{LED}$ | 0.85 | $[-]$ |
| $MW$ | 180 | $g\ mol^{-1}$ |
| $\Delta E_{photo}$ | 2807 | $kJ\ mol^{-1}$ |
| $u$ | 0.4 | $m\ s^{-1}$ |

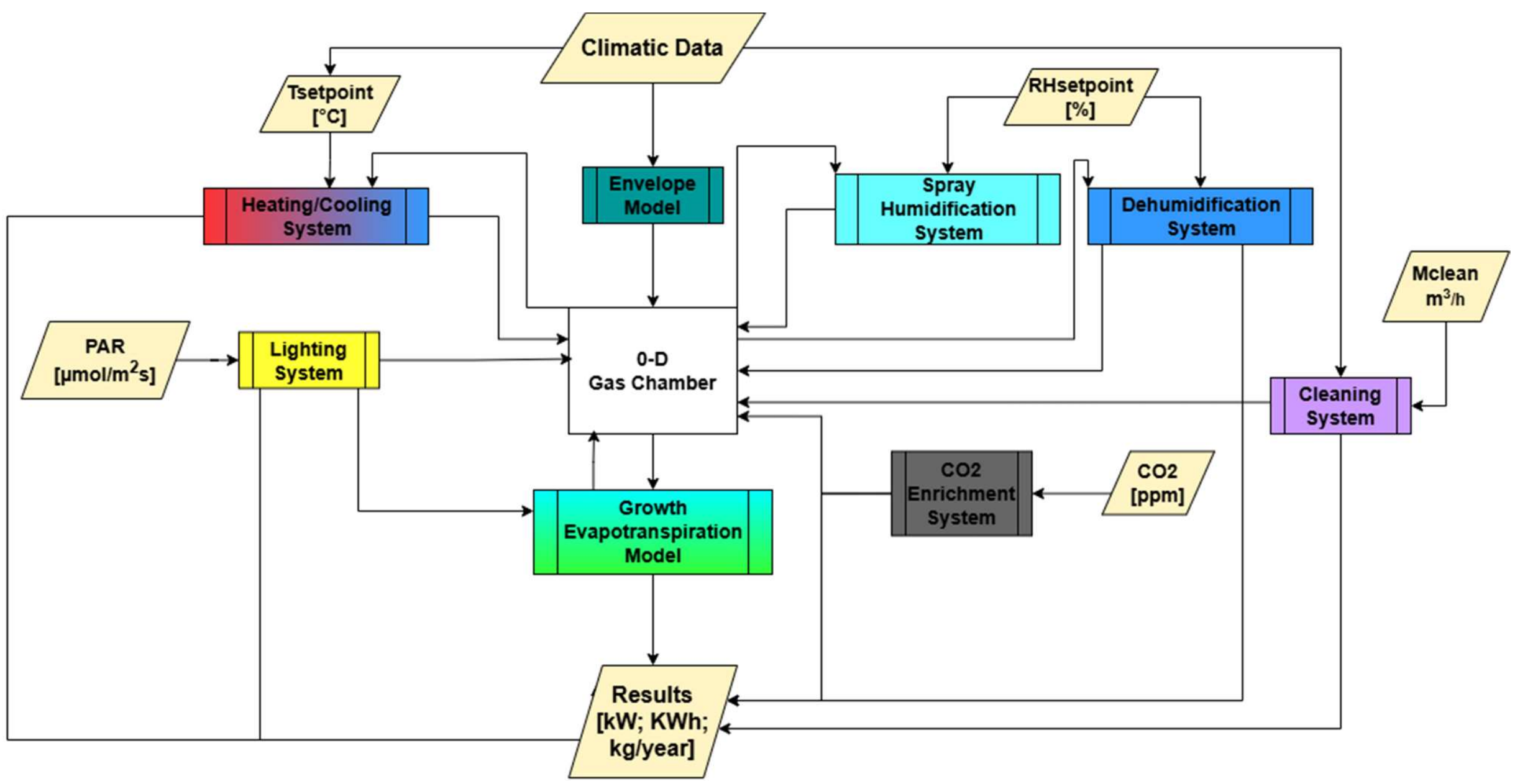

*Figure 1: Flowchart of the model developed in Amesim Simcenter [27].*

### 2.2. Studied Cases

As outlined in the Introduction, this study aims to evaluate how key environmental parameters influence the resource consumption (energy, water, and carbon dioxide), productivity, and economic feasibility of VF. The primary drivers of crop productivity in VF are indoor air temperature ($T_{in}$), PPFD level (PAR), and $CO_2$ concentration ($X_{CO2}$), which were therefore selected as the core variables under investigation. Indoor temperature not only influences plant growth directly but also impacts energy losses through the building envelope. Based on Carotti [5], who identified 24 °C as the optimal temperature for lettuce, three temperature setpoints (20 °C, 24 °C, and 28 °C) were tested to evaluate their impact on overall VF performance and energy consumption. These values represent commonly adopted temperatures in vertical farming studies. Although varying indoor temperatures inherently alter the vapor pressure deficit (VPD), this study assumes a controlled relative humidity (RH) of 75% during the photoperiod and 85% during the dark period. While RH has not shown a significant direct impact on crop growth rates [22], it does influence the energy consumption of the AHU system. Conversely, PPFD significantly affects the overall energy demand of the VF. Higher PPFD levels directly increase electricity consumption due to the lighting system and indirectly raise the HVACD load, as excess heat generated by the LEDs must be removed to maintain set indoor temperatures. On the other hand, the increase in PPFD levels leads to higher crop productivity within the vertical farm. However, this relationship is not linear, as photosynthetic efficiency declines with increasing PPFD. Carotti et al. [5] demonstrated that increasing the PPFD above 400 µmol $m^{-2}$ $s^{-1}$ is not advantageous, even from an agronomic perspective, as the resulting improvements in crop growth rate are marginal. Based on these findings, the present study explores the effect of three distinct PPFD levels on vertical farm performance: a very low level (100 µmol $m^{-2}$ $s^{-1}$), a standard reference value (250 µmol $m^{-2}$ $s^{-1}$), and a high level (400 µmol $m^{-2}$ $s^{-1}$), to assess their impact on both productivity and energy consumption. Although the positive effect of $CO_2$ enrichment on crop growth has been well established, and $CO_2$ concentration does not directly influence the energy consumption of the vertical farm, it is still relevant to quantify its impact on productivity. This allows for a meaningful comparison between the yield gains and the associated increase in operational expenditures due to carbon fertilization. For this reason, three levels of $CO_2$ concentration were tested in this study: ambient (400 ppm), moderate enrichment (900 ppm), and high enrichment (1400 ppm). The high $CO_2$ enrichment level was chosen based on studies [24] indicating that higher concentrations could negatively affect lettuce growth. Ultimately, according to [5], the optimal photoperiod for maximizing LUE, and thereby crop yield, was identified as a 16-hour light period per day. For this reason, this photoperiod was adopted in the simulations presented in the present study.

From an energy perspective, as highlighted in the current literature, vertical farming systems typically require substantial cooling loads to dissipate the thermal energy generated by LED lighting. Consequently, assessing the influence of the thermal insulation on the VF's energy consumption becomes noteworthy. In this study, two building envelope configurations were simulated: one equipped with a 105 mm thick rigid polyurethane (PUR) insulation, and one without any insulation. The impact of insulation on overall energy consumption is intrinsically linked to the external climatic conditions. To account for this, three distinct climate scenarios were selected, each representative of different latitudes and weather profiles, as well as varied socio-economic contexts. Specifically, the cities of Trondheim (cold climate), Shanghai (temperate climate), and Dubai (hot climate) were chosen. These metropolitan areas were selected not only for their contrasting climates but also because urban settings are particularly relevant for vertical farming due to land use constraints and urban food production potential. Hourly data for ambient temperature, relative humidity, and solar irradiation were obtained from [28] for the three locations. Additionally, the geographic coordinates of each city were used to compute altitude and azimuth angles, allowing for an accurate estimation of solar gains on the vertical farm's external surfaces.

All scenarios described above were simulated using the model presented in [21] and developed in Amesim Simcenter [27], assuming a square growing chamber with 7-meter sides and a height of 3 meters. One-year-long simulations were performed with outputs recorded at 15-minute intervals. The chamber was equipped with five three-tier shelving units, resulting in a net cultivated area of 90 $m^2$ (61% land use efficiency per single tier). Lettuce crops were cultivated at a density of 25 plants per square meter, under a 16-hour photoperiod, and harvested upon reaching an average fresh weight of 250 g per plant [5].

All the analyzed scenarios are summarized in Figure *2*, and from this point onward, they will be referenced using the acronyms reported therein. To evaluate the influence of each variable on the energy requirements, crop productivity, and water use efficiency (WUE) of the vertical farm, previous studies such as [18] employed the Spearman rank correlation coefficient. However, due to the non-linear relationships among the variables analyzed in this study, the distance correlation coefficient has been used instead. This metric was calculated according to Eq. 10, where dCov represents the distance covariance between two variables (X, Y), and dVar refers to their distance variance. Further details regarding the calculation procedure are provided in the Appendix. A distance correlation coefficient (ρ) approaching 1 indicates a strong correlation between the variable and the outcome, while a value of ρ near zero suggests a negligible or no correlation.

$$\rho(X,Y) = \frac{dCov(X,Y)}{\sqrt{dVar(X) \cdot dVar(Y)}} \quad (10)$$

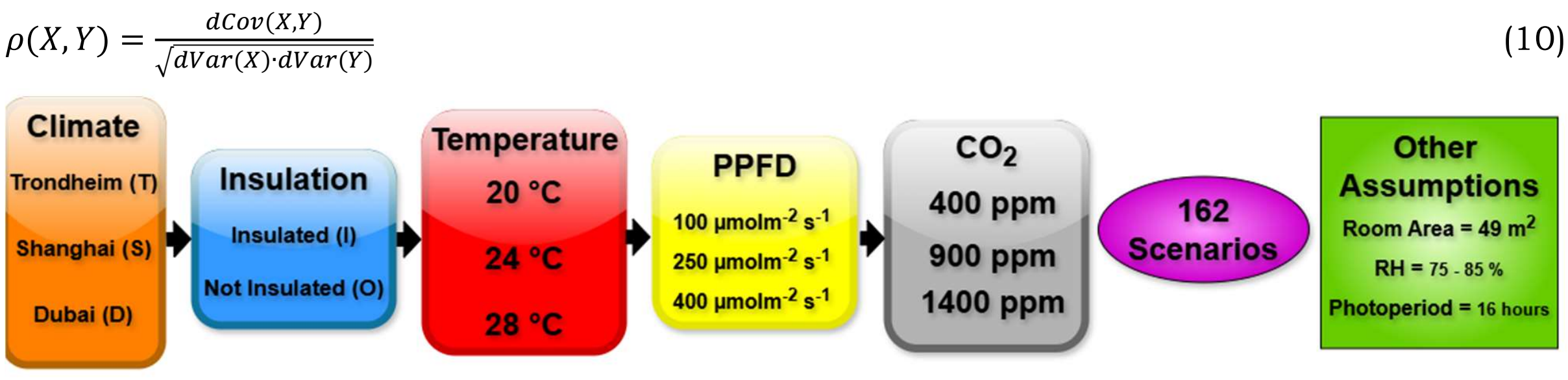

Figure 2: *Input conditions for the scenarios assessed by sensitivity analysis.*

### 2.3. Economic and Sustainability Analysis

The feasibility of a VF is significantly influenced by its location. Therefore, this section outlines all assumptions related to the costs and emissions associated with lettuce production in vertical farming systems. In line to decarbonize the agricultural sector through electrification [23], two air-to-water heat pumps (HPs) were employed to meet the energy demands for heating, cooling, and dehumidification within the HVACD system. The electricity consumption of the HPs was defined in Eq. (11). The coefficient of performance (COP) was calculated using an algorithm that considers for the nominal COP ($COP_{nom}$) (which depends on the HP size), external temperature, indoor temperature and part-load factor (PLF) as described in Eq. (12). For the HP used for heating purposes, the dependencies of the COP on temperature ($COP_T$) and part-load factor were sourced from [29] and [30], respectively. For the unit used for cooling and dehumidification, data were obtained from [31]. This method is deemed more accurate than the exergy efficiency assumptions reported in [21]. The nominal COP values for the HPs were determined based on commercial data provided in the AERMEC catalogue [32], considering the appropriate duty. Further methodological details can be found in [21].

$$P_{HP}(t) = \frac{Q_{HP}(t)}{COP(t)} + 0.008 \cdot Q_{HP}(t) \quad (11)$$

$$COP(t) = COP_{nom} \cdot \frac{COP_T(T_{ext}, T_{wat})}{COP_T(T_{ext,nom}, T_{wat})} \cdot PLF(PLR) \quad (12)$$

The first step in evaluating the feasibility of the VF involved assessing the capital expenditures (CAPEX) required for its implementation. Assuming the building structure is already in place, the fixed expenditures include the lighting system, electrical wiring, HVACD system, growing chamber, and insulation installation. These assumptions were reported in Table *2*. These costs were considered location-independent, and a linear scaling function was applied to estimate the total investment. The specific costs for lighting, wiring, growing, and HVACD, sourced from [33], were adjusted (+27%) using the Consumer Price Index (CPI) trend from 2018 to 2025, as reported by the U.S. Bureau of Labor Statistics [34]. The LED cost per

W reported in [33] was based on an LED efficiency of 1.48 μmol $m^{-2}$ $s^{-1}$; therefore, this value was doubled to reflect the higher LED efficiency adopted in this study. Additionally, the purchasing cost of the polyurethane rigid foam (PUR) used for insulation was increased by 40% to account for installation expenses [35].

Table 2: A*ssumed costs for vertical farm purchase [33], [35].*

| Symbol | Value | Adjusted Value | Unit |
|---|---|---|---|
| $c_{Light}$ | 1.48 | 3.76 | $\$\,W^{-1}$ |
| $c_{Wir}$ | 0.85 | 1.08 | $\$\,W^{-1}$ |
| $c_{HVACD}$ | 1.092 | 1.39 | $\$\,W^{-1}$ |
| $c_{VF}$ | 224 | 284.5 | $\$\,{m_{crop}}^{-2}$ |
| $c_{Ins}$ | 20 | 28 | $\$\,m^{-2}$ |

Since all scenarios were analyzed and compared over a year-long simulation, it is essential to estimate the annual operational expenditures (OPEX) associated with lettuce production. Table *3* summarizes the reference prices adopted in this section. The OPEX includes crop-related costs (seeds, nutrients, and packaging), expressed in $ per kilogram of lettuce produced, as well as costs for electricity, water, carbon dioxide, labor, and land leasing. Crop-related costs were assumed to be location-independent. Conversely, electricity, water, labor, and leasing prices were evaluated based on the specific conditions of each study location. In recent years, the price of carbon dioxide has risen, partly due to increased demand for carbon fertilization [36]. Therefore, an average global price of 3.5 $ $kg^{-1}$ was adopted. Labor costs were estimated under the assumption that workers are trained agronomists, with a labor requirement of 0.067 hours per kilogram of lettuce produced [37]. Lastly, land leasing costs were assessed based on average rental prices of warehouses in the selected cities.

Table 3: *Assumed costs for estimating operational expenditures.*

| Symbol | Trondheim | Shanghai | Dubai | Unit |
|---|---|---|---|---|
| $c_{crop}$ | 1.14 [33] | 1.14 [33] | 1.14 [33] | $\$\,kg^{-1}$ |
| $c_{CO2}$ | 3.50 [38][39] | 3.50 [38][39] | 3.50 [38][39] | $\$\,kg^{-1}$ |
| $c_{el}$ | 0.10[40] | 0.09[41] | 0.11[42] | $\$\,kWh^{-1}$ |
| $c_{wat}$ | 2.90[43] | 0.48[41] | 2.51[42] | $\$\,m^{-3}$ |
| $c_{Lab}$[1] | 33.33[44] | 14.43[44] | 30.93[44] | $\$\,h^{-1}$ |
| $c_{Leas}$ | 143.40 [45] | 50.10[46] | 168.50 [47] | $\$\,m^{-2}$ |

The aforementioned assumptions have been applied to evaluate both the capital expenditure (CAPEX) and the operational expenditure (OPEX) of the studied vertical farms using Equations 13 and 14. Since the key parameter for the economic assessment is the levelized cost of lettuce (LCoL), an average system lifetime (n) of 20 years [48] and a nominal interest rate ($r_{nom}$) of 8.5 percent, as typically assumed for agricultural investments [49] have been considered. Assuming an inflation rate of 2%, the real interest rate ($r_{real}$) has been calculated (see Eq. 15). This rate has been used to determine the capital recovery factor (CRF), which is then applied to annualize both the initial investment (CAPEX) and the replacement costs. Given that the estimated lifespan of the LEDs is shorter than the overall system lifetime of 20 years, the cost of replacing the lighting system ($C_{Rep}$) has been included twice: once after 8 years and once again after 16 years from the start of operations. These future expenses have been converted to their present value using the real interest rate, as described in Equation 16. Finally, the sum of the annualized investment and replacement costs, along with the annual OPEX, has been divided by the annual crop yield of the VF to calculate the LCoL, as outlined in Equation 18.

[1] The referenced source provides the comprehensive dataset used to determine the average hourly wage for each location.

$$CAPEX = c_{Light} \cdot P_{Light} + c_{Wir} \cdot (P_{Light} + P_{HVACD}) + c_{HVACD} \cdot P_{HVACD} + c_{VF} \cdot A_c + c_{Ins} \cdot A_{env} \tag{13}$$

$$OPEX = M_{crop} \cdot c_{crop} + E_{el} \cdot c_{el} + M_{wat} \cdot c_{wat} + M_{CO2} \cdot c_{CO2} + h_{lab} \cdot c_{lab} + A_{VF} \cdot c_{leas} \tag{14}$$

$$r_{real} = \frac{(1-r_{nom})}{(1-i)} - 1 \tag{15}$$

$$CRF = \frac{r_{real} \cdot (1+r_{real})^n}{(1+r_{real})^n - 1} \tag{16}$$

$$C_{Rep} = (c_{Light} \cdot P_{Light}) \cdot \left(\frac{1}{(1+r_{real})^8} + \frac{1}{(1+r_{real})^{16}}\right) \tag{17}$$

$$LCoL = \frac{(CAPEX + C_{Rep}) \cdot CRF + OPEX}{M_{crop}} \tag{18}$$

Ultimately, the energy mix of the selected locations, Trondheim [50], Shanghai [51], and Dubai [52] was used to estimate the carbon emissions associated with vertically grown lettuce. To provide a benchmark for lettuce-related carbon emissions, transport-related emissions from imported lettuce were also calculated for each location, based on the assumptions listed in Table *4*. The supply chain for each location and the corresponding mode of transport (refrigerated truck, refrigerated ship, or airplane) is chosen according to the supply distance. The exporting country for each location was identified using data provided by the World Integrated Trade Solution for Norway [53], China [54], and United Arab Emirates [55]. In this study, truck transportation is assumed to be conducted using a 32-ton articulated vehicle, with an average payload of 9.33 t (corresponding to 90% of the total capacity) and an average volume load of 14.38 pallets. To estimate specific carbon emissions, chilled single-drop distribution is considered, along with an average annual refrigerant leakage of 15% R404A [56]. Assuming a lettuce density of 200 kg $m^{-3}$ and an available refrigerated container volume of 27 $m^3$ (90% of 1 twenty-foot equivalent unit (TEU)), emission data corresponding to average speed and refrigerated containers, as reported in [57], were used. Airplane-related emissions are assumed to range from 4 to 10 times higher than those of maritime transport [58]; therefore, an average multiplier of 7 was applied to estimate emissions from air freight.

Table 4: *Assumptions for estimating the difference in $CO_2$ emissions between local and imported lettuce.*

| **Symbol** | **Trondheim** | **Shanghai** | **Dubai** | **Unit** |
|---|---|---|---|---|
| Export Country | Spain | Korea | Spain | – |
| $L_{supply}$ | 3400 | 1000 | 5000 | $km$ |
| Transport | Truck (100%) | Truck (20%) + Ship (80%) | Truck (5%) + Airplane (95%) | – |
| $e_{VF}$ | 15 | 585 | 404 | $ton_{CO2-eq}/GWh$ |
| $e_{truck}$ | 98.30 | 98.30 | 98.30 | $g_{CO2-eq}/(ton \cdot km)$ |
| $e_{ship}$ | 35.71 | 35.71 | 35.71 | $g_{CO2-eq}/(ton \cdot km)$ |
| $e_{airplane}$ | 250.00 | 250.00 | 250.00 | $g_{CO2-eq}/(ton \cdot km)$ |

### 2.4 Analysis Uncertainties and Limitations

The Agri-Energy model used in this study has been validated with experimental data and through intermodel comparison. As highlighted in [21], the lettuce growth submodel achieved an average $R^2$ of 0.97 against experimental data from [5]. Energy validation via comparison with Talbot et al. [22] showed a yearly energy load offset below 8%. The economic submodel is based on average costs collected from the literature, and a relative weight analysis, reported in the Appendix, was included to assess the impact of individual costs on CAPEX and OPEX.

The main limitation of this analysis lies in the crop growth and evapotranspiration submodel, which is specific to lettuce. The lack of detailed experimental data for other crops prevents extending the analysis to additional species. The vertical farm is represented by a single architecture, featuring a two-heat-pump, grid-powered HVACD system. While this choice reflects the most common VF configuration, it does not provide insights into the impact of renewable energy integration on VF performance and cost-effectiveness.

## 3. Results and Discussion

In this section, the results of the comprehensive analysis of the cases studied are presented. The chapter is organized into eight sections addressing crop growth, the thermal demand of the VF, the achieved water use efficiency, the electricity requirement, and the economic effectiveness and sustainability of the studied VFs. Finally, two subsections highlight the results of the distance correlation analysis between inputs and outputs.

### 3.1. Crop Growth Results

As reported in related studies, indoor temperature, PPFD, and $CO_2$ concentration are critical determinants of lettuce growth in VF systems. Figure *3* illustrates their contributions to the growth cycle (from transplanting to harvest). The data confirm that the effect of $CO_2$ fertilization diminishes as PPFD increases: at 100 μmol $m^{-2}$ $s^{-1}$, elevated $CO_2$ shortened the cycle by 31% compared to ambient conditions (400 ppm), whereas at 400 μmol $m^{-2}$ $s^{-1}$ the reduction decreased to 22%. Moreover, $CO_2$ enrichment above 900 ppm provided only marginal benefits and, under high PPFD, was occasionally associated with slight performance declines.

The effect of PPFD is stronger at lower indoor temperatures, but its impact on growth rate is non-linear since LUE also varies with light intensity. Above 250 μmol $m^{-2}$ $s^{-1}$, LUE decreases significantly, producing the asymptotic trend shown in the figure. Consequently, for any given indoor temperature and $CO_2$ concentration, the reduction in the growth cycle length with increasing PPFD progressively diminishes at higher PPFD levels.

The impact of temperature on crop growth was relatively limited. Increasing indoor temperature to 24°C and 28°C shortened the growth cycle by 17% and 14%, respectively. However, higher temperatures reduced LUE, thereby attenuating the positive effect of higher PPFD. At 20°C, raising PPFD to 400 μmol $m^{-2}$ $s^{-1}$ reduced the growth cycle by 20.5%, whereas at 24°C and 28°C the reduction improved to 18.4% and 7%. Overall, annual productivity ranged from 31 to 118 kg $m^{-2}$, corresponding to the output of 550-2100 $m^2$ of open-field cultivation. This translates into a land use efficiency 11-43 times greater than conventional open-field farming.

A key advantage of using a detailed lettuce growth model is the improved accuracy in yield estimation compared to previous studies. For instance, under the same PPFD and temperature conditions (250 μmol m$^{-2}$ s$^{-1}$, 400 μmol m$^{-2}$ s$^{-1}$, and 24°C), [18] reported fixed yields of 75 and 108.8 kg m$^{-2}$, respectively, whereas the adopted model, which accounts for carbon fertilization, estimated ranges of 75-102 kg m$^{-2}$ and 92-118 kg m$^{-2}$. Similarly, the economic analysis by [17] assumed a constant yield of 150 kg per land square meter, while the detailed analysis provided a broader and more realistic range of 57-217 kg per land square meter (-62%, +45%).

In conclusion, for the crop production perspective, the optimal indoor conditions within the growing chamber are an indoor temperature of 24°C, a PPFD level of 400 μmol m$^{-2}$ s$^{-1}$, and a $CO_2$ concentration of 900 ppm, resulting in a crop growth cycle of only 19 days. This result is consistent with the findings reported in [5], where the highest productivity was achieved at the highest PPFD. While PPFD levels above 400 μmol m$^{-2}$ s$^{-1}$ might further reduce the growth cycle duration, such scenarios were not considered in this study, as previous research has shown that the additional energy demand would outweigh the production benefits.

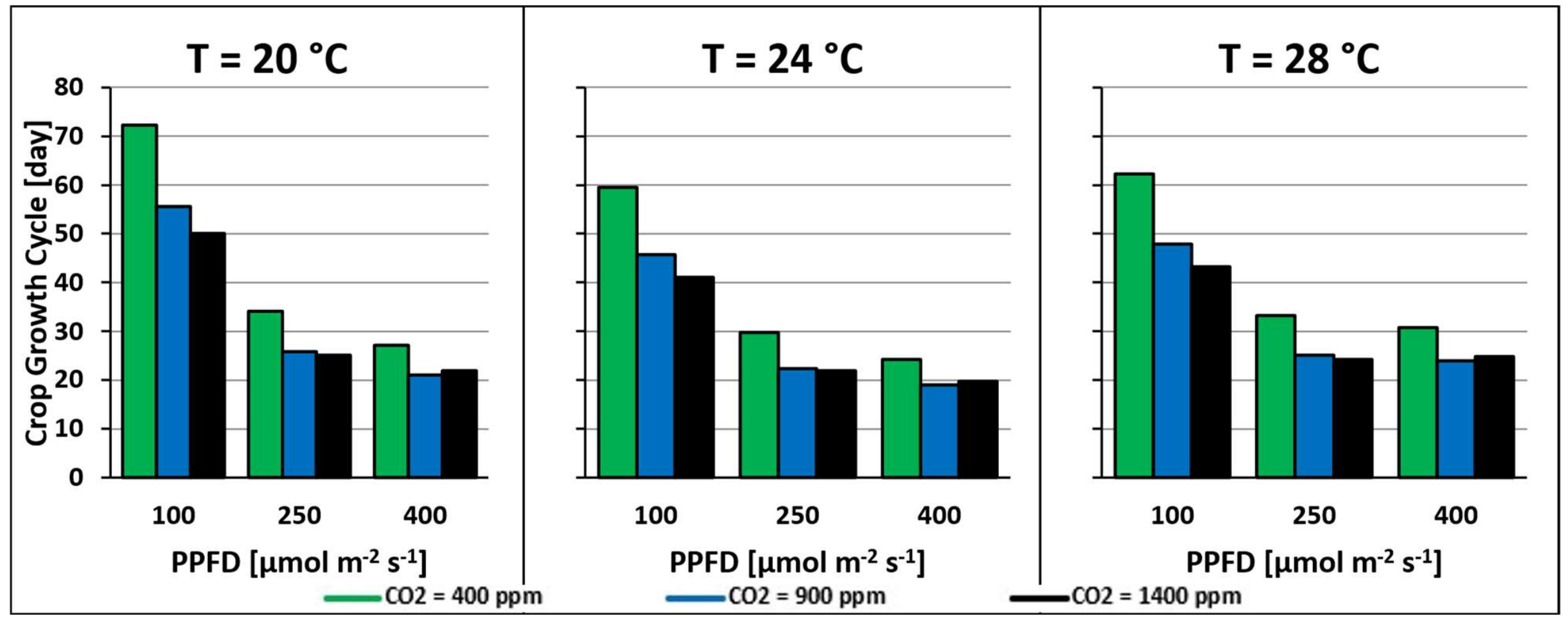

Figure 3: *Effect of indoor condition (Temperature, PPFD, and $CO_2$ concentration) on crop growth cycle length.*

### 3.2. Vertical Farm Thermal Demand

The thermal demand of a VF includes heating to maintain suitable temperatures during dark periods, cooling to remove excess heat from lighting, and dehumidification with post-heating (PH) to manage plant evapotranspiration. These energy fluxes are driven by indoor temperature, PPFD, and external climate. $CO_2$ concentration also affects energy demand through its impact on crop growth. However, as its impact on annual loads is minor (<2%), it will not be explicitly considered in the energy demand analysis.

Among the thermal energy fluxes, cooling is the dominant load, as shown in Figure *4* for an insulated VF operated at 20 °C and 100 μmol m$^{-2}$ s$^{-1}$, based on year-round simulations in the three locations. Cooling also contributes to moisture control by condensing water vapor, which reduces dehumidification requirements. Both loads vary with crop development: in early stages, limited evapotranspiration and small leaf area often require spray humidification to compensate for moisture removal by cooling. This process provides a "free cooling" effect, lowering net cooling demand. In later stages, evapotranspiration rises sharply, increasing dehumidification needs while eliminating the free cooling effect, which raises the cooling load again. These dynamics generate the characteristic wavy cooling trend observed in Figure *4* and Figures A2–A9 (Appendix).

Figures A1–A8 present the thermal load trends for all simulated scenarios and are included in the Appendix to improve the manuscript's readability and reduce its length. Despite being placed in the supplementary section, these figures clearly illustrate the effects of the three investigated parameters (indoor temperature, PPFD, and geographic location) on the required thermal loads.

As previously discussed, PPFD strongly influences both cooling and dehumidification loads. Higher light intensity increases the heat to be removed from the growing chamber, but the rise in cooling demand is non-linear with PPFD due to LUE variability. Raising PPFD from 100 to 250 and 400 $\mu mol \cdot m^{-2} \cdot s^{-1}$ increases the cooling load by 180% and 360%, respectively. Conversely, dehumidification demand decreases significantly (- 67% and - 77%) because the cooling system condenses much of the moisture. At PPFD levels above 100 $\mu mol\ m^{-2}\ s^{-1}$, dehumidification is required mainly during dark periods, when cooling is inactive. Heating demand, conversely, occurs almost exclusively in dark periods and is largely unaffected by PPFD.

Thanks to the insulation, the effect of indoor temperature on cooling demand is limited: increasing temperature from 20 °C to 28 °C reduces cooling load by 8.5%-10%, depending on the external climate. Conversely, heating demand remains strongly climate-dependent, rising by 50% in cold climates (e.g., Trondheim) and up to 360% in hot climates (e.g., Dubai) for the same temperature increase (from 20°C to 28°C). Higher indoor temperatures also increase the allowable vapor pressure deficit (VPD) by approximately 60%, reducing the dehumidification load by 48% and 58% in cold and hot climates, respectively.

Consequently, the impact of geographic location on an insulated vertical farm is relatively limited, as illustrated in Figure *4*. However, humid climates can slightly increase the dehumidification load due to the higher moisture content of the outdoor air used by the air exchange system, which is necessary to remove potentially harmful compounds produced during photosynthetic reactions. In conclusion, among all the tested scenarios, the configuration identified as T_I_100_28_900, characterized by a cold climate, low PPFD, and high indoor temperature, resulted in the lowest overall thermal energy demand, with an annual thermal load of 18 $MWh_{th}$.

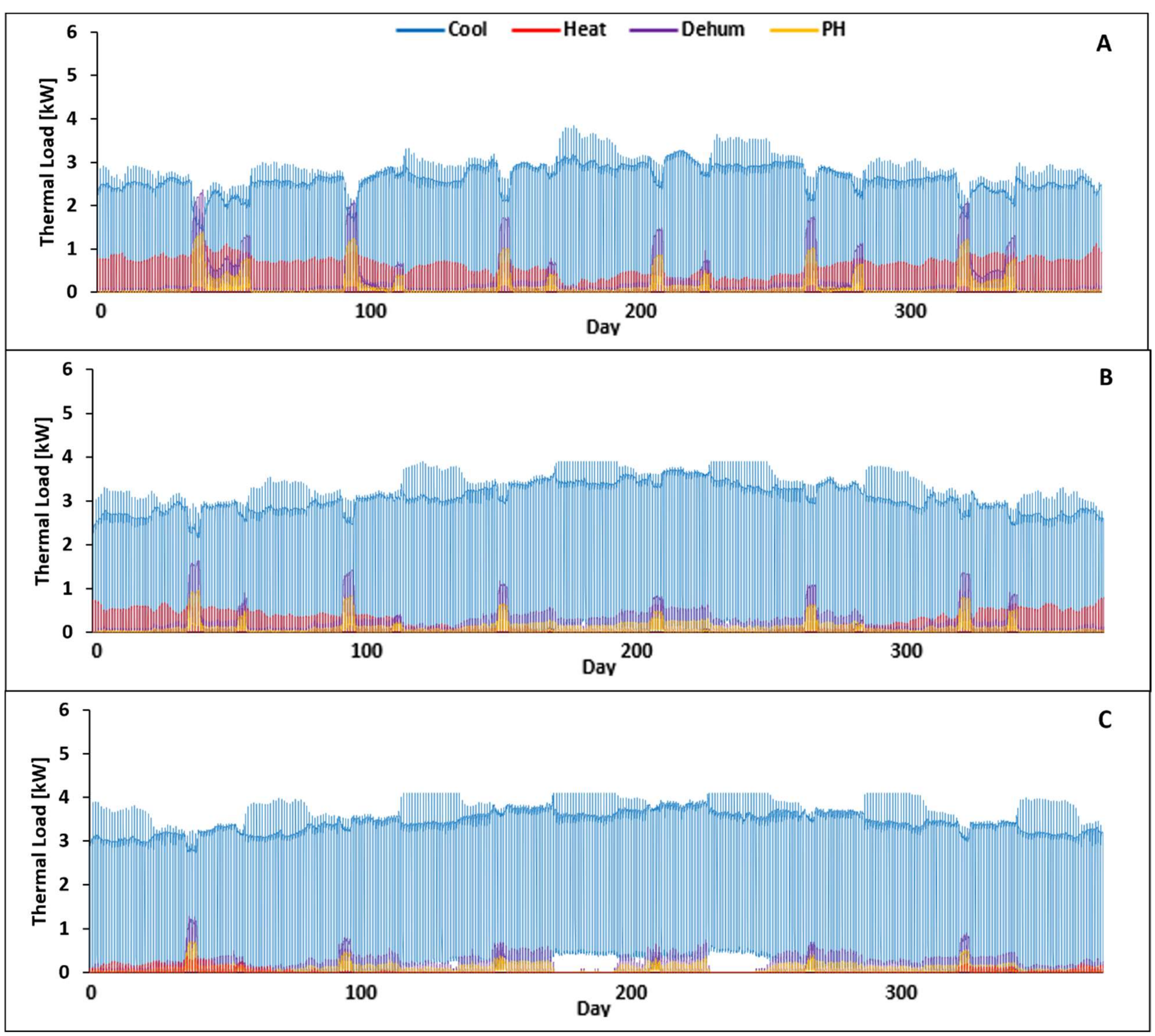

Figure 4: *Heating, cooling, dehumidification, and post-heating loads throughout the year for a well-insulated VF in three locations: (A) Trondheim, (B) Shanghai, (C) Dubai. The results were obtained under conditions of 20 °C and 100 μmol m$^{-2}$ s$^{-1}$.*

The following paragraphs illustrate the impact of insulation on thermal loads across the studied locations. In insulated VFs, heating and cooling are relatively independent of external climate, whereas Figure *5* (based on the same scenario of Figure *4*) shows that without insulation, these loads are significantly climate-dependant. Full results for all scenarios are provided in Figures A10–A17 (Appendix). Without insulation, heating and cooling exhibit opposite trends: in cold climates under low PPFD. The heating system operates even during light periods, while in hot climates, the cooling system remains active during dark periods to offset heat gains through the building envelope. Dehumidification demand increases when cooling demand decreases, primarily due to heat losses, as typically observed in cold climates.

In cold climates with low PPFD and indoor temperature, removing insulation increases the heating load by up to 360%, and reduces the cooling demand by approximately 70%. However, this reduction in cooling load leads to an increase in both dehumidification and post-heating loads, resulting in an overall thermal energy increase of about 40%. This effect intensifies at higher indoor temperatures due to greater heat losses through the building envelope: at 28 °C and the same PPFD, total energy consumption increases by 75% compared to the insulated case. Higher PPFD partially mitigates this impact: at 400 μmol $m^{-2}$ $s^{-1}$ and 20 °C, the heating load increases by 280%, cooling and dehumidification loads decrease by around 11%, yielding an overall energy reduction of about 4%, rising to 5% at higher temperatures.

In hot climate conditions, removing the insulation has a limited impact on the heating load, as most of the thermal gains come from the lighting system. At 20°C (indoor temperature), the heating load increases by approximately 107%, while the cooling load rises by 7–27%, depending on PPFD. Higher PPFD reduces the relative effect of insulation on total energy consumption. At low PPFD, the overall thermal energy consumption increases 6-21%, substantially less than in cold climates. Nonetheless, insulation removal consistently raises overall thermal energy demand. Among the tested scenarios, the one most resilient to the absence of insulation is characterized by an indoor temperature of 28 °C and a PPFD of 400 μmol $m^{-2}$ $s^{-1}$.

In conclusion, the lowest total energy consumption among the non-insulated VF scenarios was observed in configuration D_O_100_28_900, with an annual thermal energy demand of 21 $MWh_{th}$. However, the removal of the insulation can only be considered a viable strategy under specific conditions, namely, in colder climates where both the indoor temperature and PPFD are maintained at high levels.

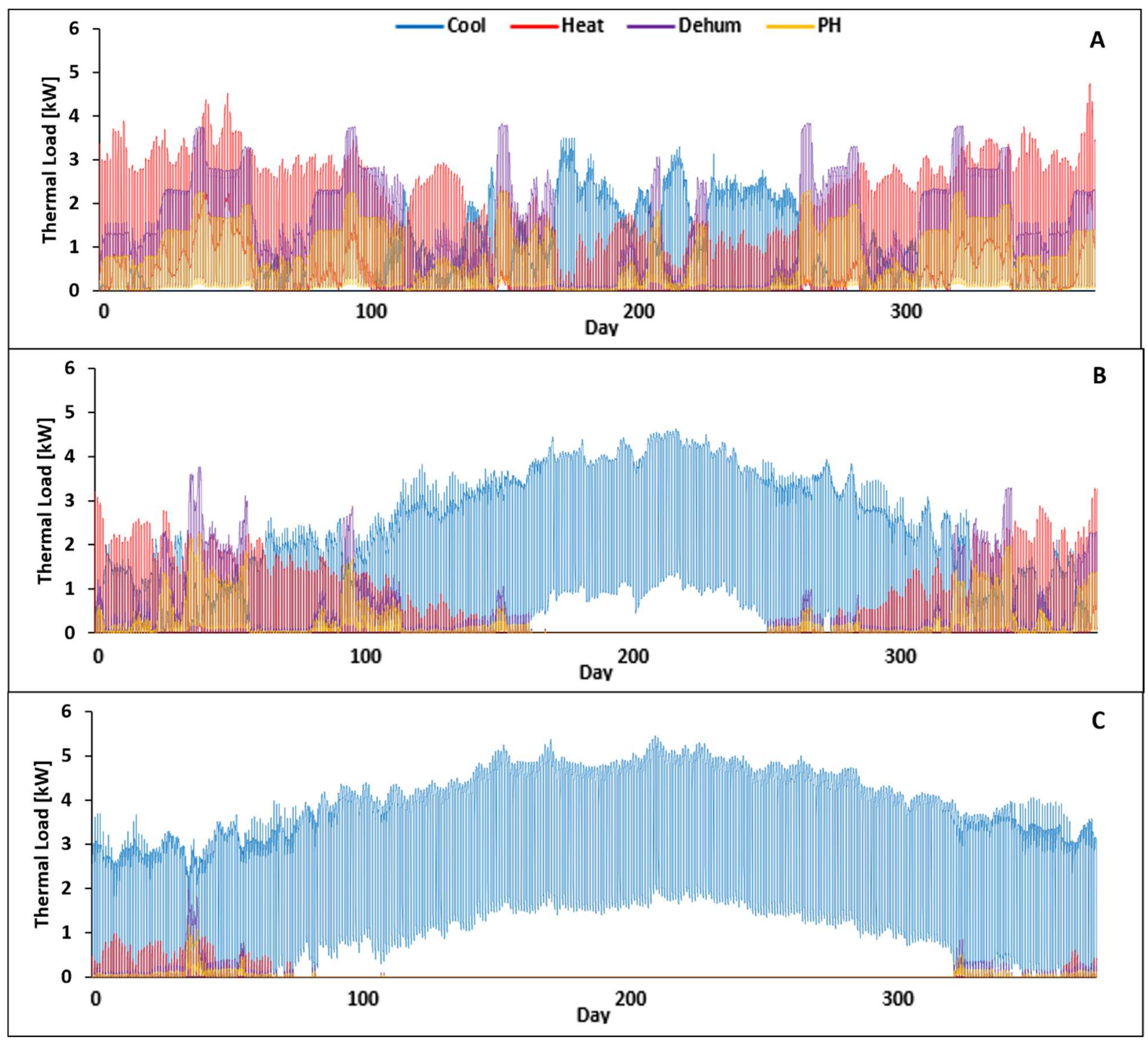

Figure 5: *Heating, cooling, dehumidification, and post-heating loads throughout the year for a not-insulated VF in three locations: (A) Trondheim, (B) Shanghai, (C) Dubai. The results were obtained under conditions of 20 °C and 100 µmol $m^{-2}$ $s^{-1}$.*

### 3.3. Specific Energy Consumption and Water Use Efficiency

As highlighted in the previous section, the insulation plays a modest role in the VF's energy performance. Although the absence of insulation may reduce energy consumption under conditions of high indoor temperature and high PPFD, i.e., when the crop's LUE is lower, insulating the VF remains preferable from an energy efficiency standpoint in hot climates, or under low to medium PPFD levels. Therefore, only the specific energy consumption (SEC) of the simulated insulated scenarios is presented in Figure *6*, while the corresponding SEC values for non-insulated scenarios are provided separately in Figure A17 in the Appendix.

In all simulated scenarios where the VF is properly insulated, it is evident that the specific energy demand (SEC) is primarily driven by the PPFD. In fact, LED lighting and the associated cooling load required to dissipate the heat generated by the lighting system account for over 95% of the total SEC, regardless of the indoor temperature or geographic location. Overall, the SEC ranges from 7.8 to 23.8 kWh kg$^{-1}$, with the lowest values observed in scenarios combining low PPFD, high $CO_2$ concentration, and an indoor temperature of 24 °C. Increasing the indoor temperature from 20 °C to 24 °C results in an SEC reduction of 11–24%, depending on the PPFD level, with the greatest savings occurring at the lowest PPFD. At 28 °C, further energy savings are marginal for low PPFD scenarios, while at 400 μmol m$^{-2}$ s$^{-1}$, raising the temperature to 28 °C leads to a 13% increase in SEC. For PPFD levels of 100 or 250 μmol m$^{-2}$ s$^{-1}$, higher $CO_2$ concentrations correlate with lower SEC, due to the enhanced crop growth rates. However, at higher PPFD levels, the optimal $CO_2$ concentration appears to be 900 ppm, beyond which the benefits plateau or diminish. Ultimately, scenarios with lower PPFD result in the lowest SEC. According to the experimental data reported by Carotti et al., this outcome is attributable to the increase in light use efficiency (LUE) and specific leaf area (SLA) as PPFD decreases. Both factors contribute to faster crop growth, thereby enhancing productivity relative to the amount of incoming light energy. However, this improvement in energy efficiency comes at the cost of reduced crop productivity per unit area, which drops from 118 kg m$^{-2}$ (at 400 μmol m$^{-2}$ s$^{-1}$) to 55 kg m$^{-2}$ (at 100 μmol m$^{-2}$ s$^{-1}$). Nevertheless, from a strictly energy consumption standpoint, the most favourable configuration is T_I_100_24_1400.

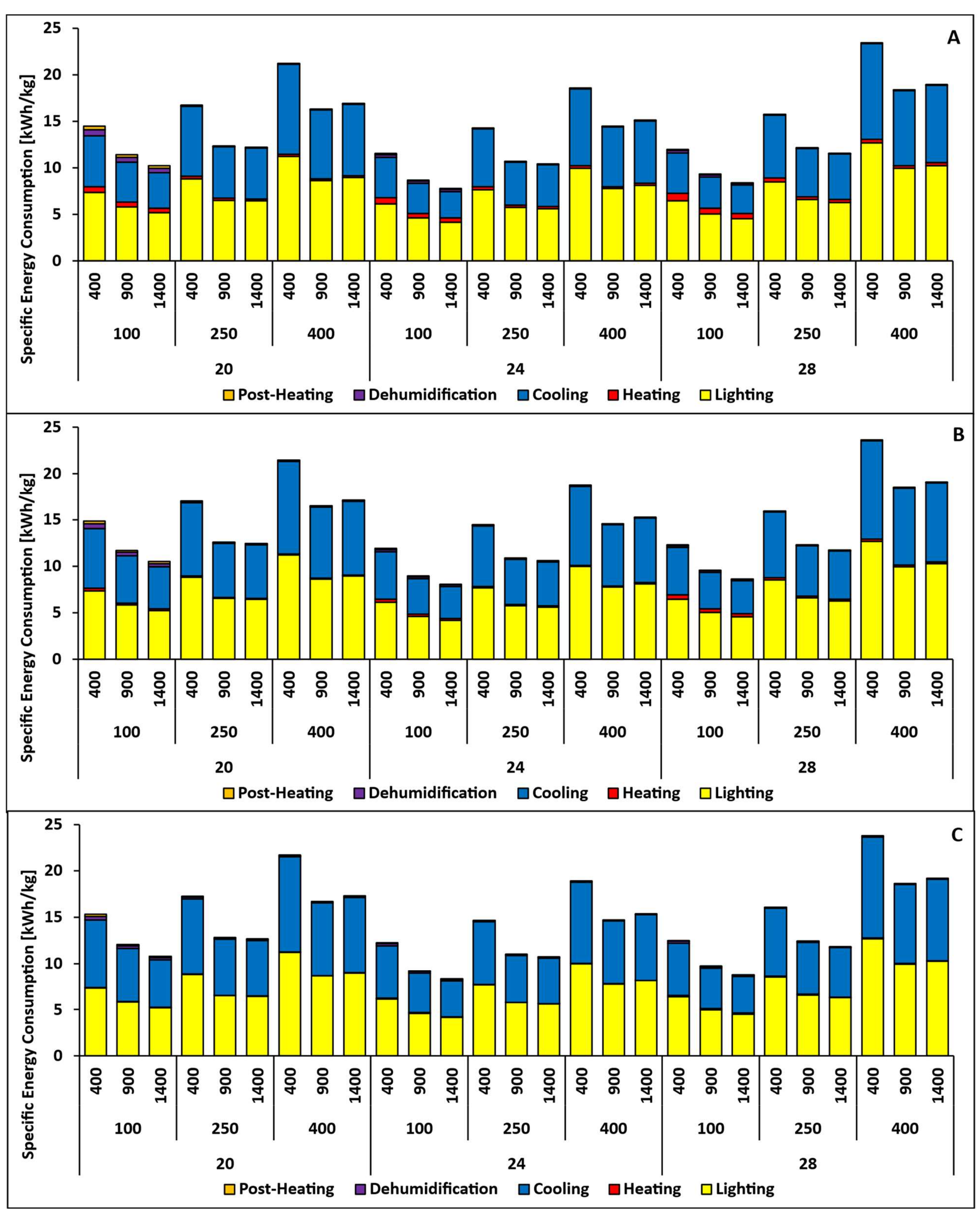

Figure 6: *Specific Energy Consumption of the VF under the three simulated locations: (A) Trondheim, (B) Shanghai, (C) Dubai.*

Water consumption is a key performance indicator for vertical farming systems. A major advantage of VF over open-field or greenhouse cultivation is the near-complete recovery of water. Water use efficiency (WUE) is defined as the ratio of total lettuce fresh weight to net water consumption of the system, where net water use equals total supplied water (crop uptake plus humidification) minus water recovered through cooling and dehumidification. Figure *7* presents the contributions of each water input and the corresponding WUE for all simulated scenarios in the selected locations. Since the differences in water requirements between different climates remain below 2%, only the Trondheim values for this parameter have been reported to improve the readability of the paper. However, the water use efficiency values for all locations are illustrated in the same figure. Specifically, the water absorbed by the crops consists of the sum of the evapotranspiration flux (ET) and the water content retained within the harvested lettuce (Crop).

The indoor temperature influences WUE by slightly reducing its value as the temperature increases. This effect is explained by the fact that the increase in crop growth associated with higher indoor temperatures is not sufficient to offset the additional water required by the system. In particular, the humidification flow (Hum) is the most affected component, showing increases of 68% and 82% when the indoor temperature rises from 20°C to 24°C and from 24°C to 28°C, respectively.

Light intensity influences WUE by increasing all water fluxes in the VF. The humidification system is the most affected, as the water required to offset moisture condensed and removed by the cooling system is directly proportional to PPFD. Higher PPFD increases cooling demand, thereby enhancing water removal. In cold climates, increasing the PPFD from 100 to 400 μmol $m^{-2}$ $s^{-1}$ improves WUE by 6-15%, depending on the indoor temperature. Conversely, in hot climates, the effect is less pronounced, while at an indoor temperature of 20°C, the trend is reversed, with WUE decreasing as the PPFD increases.

The effect of carbon dioxide concentration on WUE is relatively limited. Higher $CO_2$ levels accelerate crop growth, resulting in only minor changes in WUE. These variations are primarily influenced by external temperature and relative humidity, which affect both heat losses through the building envelope and the moisture content of the air used for cleaning purposes.

Ultimately, external climate significantly affects VF water dynamics, even when the vertical farm is well-insulated. While total water demand remains largely unchanged, higher outdoor humidity increases condensation in the chiller and the AHU, boosting water recovery. As a result, the volume of recovered water increases, allowing the WUE to exceed its theoretical maximum (defined as 1000 plus the dry matter content, DMC). Consequently, in a hot and humid climate, net water requirement decreases notably, especially at lower indoor temperatures. Generally, this translates into lower water consumption and improved WUE, although some of the recovered water is inevitably lost during indoor air exchanges. Under extreme conditions, up to 240 kg of water per year can be extracted from the incoming air, reducing the VF's net annual water requirement by around 10%. Therefore, from a water use efficiency perspective, the configuration identified as D_I_100_20_400 proved to be the most optimal, achieving a WUE of 1132 g $L^{-1}$.

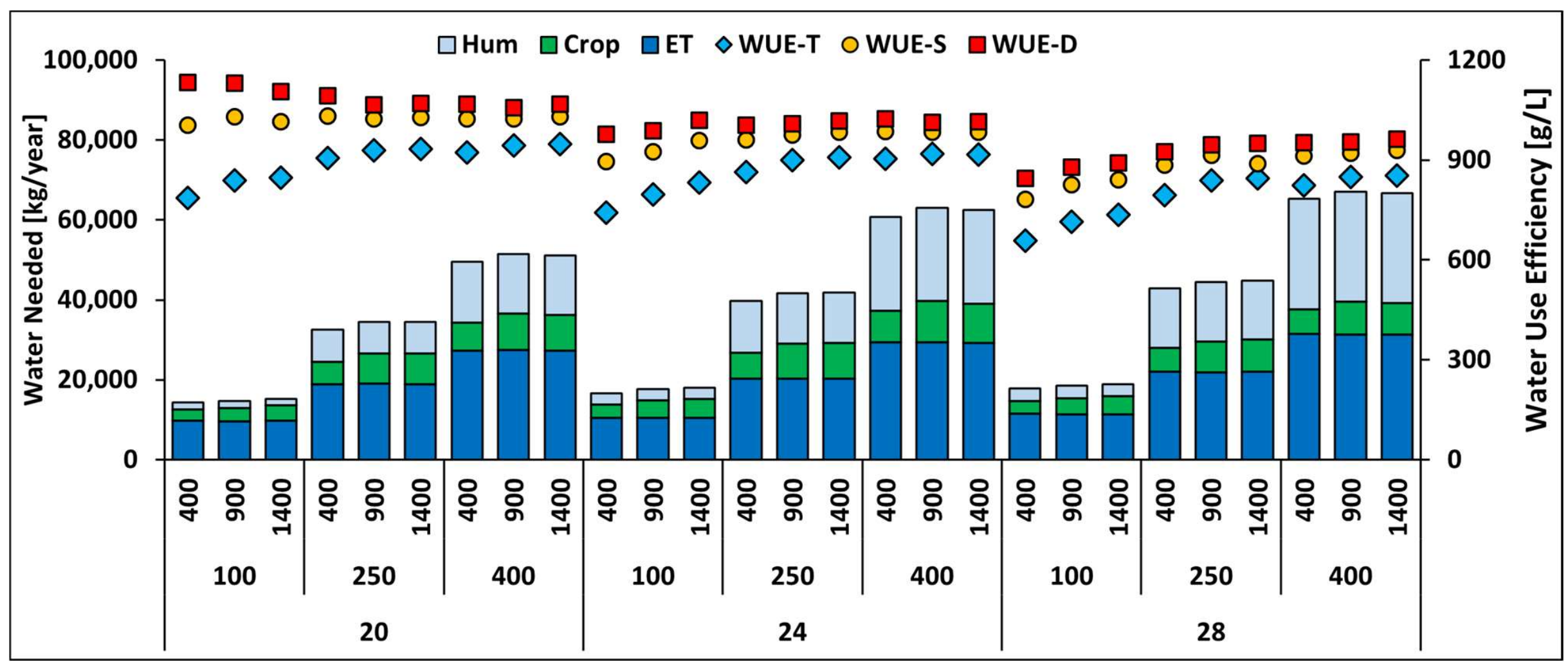

Figure 7: *Annual Water requirement and Water Use Efficiency achieved by VF under different indoor conditions in the three locations: (A) Trondheim, (B) Shanghai, (C) Dubai.*

### 3.4. Distance Correlation Coefficient Analysis

This section evaluates how each input variable influences the main outcomes and performance indicators of the vertical farm. Figure *8* presents the distance correlation coefficients ($\rho$) between indoor temperature, PPFD level, $CO_2$ concentration, insulation, and external climate, with crop production, water use efficiency, and specific energy consumption.

PPFD emerges as the most influential parameter, strongly correlated with crop production ($\rho = 0.85$), and SEC ($\rho = 0.73$, rising to 0.99 for total energy). Indoor temperature shows a limited impact on yield ($\rho = 0.18$) and SEC ($\rho = 0.22$), but a stronger link with WUE ($\rho = 0.59$) due to its effect on VPD control. $CO_2$ concentration moderately influences production ($\rho = 0.36$) and indirectly SEC ($\rho = 0.44$). Insulation has marginal effects on all indicators ($\rho \leq 0.08$), while external climate primarily impacts WUE ($\rho = 0.62$) via air humidity conditions.

From a practical perspective, these findings suggest that entrepreneurs should prioritize investment in efficient lighting systems and $CO_2$ enrichment strategies, as these most directly drive productivity and energy performance. At the same time, policymakers could incentivize innovations that improve lighting efficiency and $CO_2$ utilization, given their disproportionate impact on sustainability. Furthermore, the strong role of external climate on WUE highlights the relevance of location-specific strategies and possible policy support for siting vertical farms where air humidity profiles minimize water consumption.

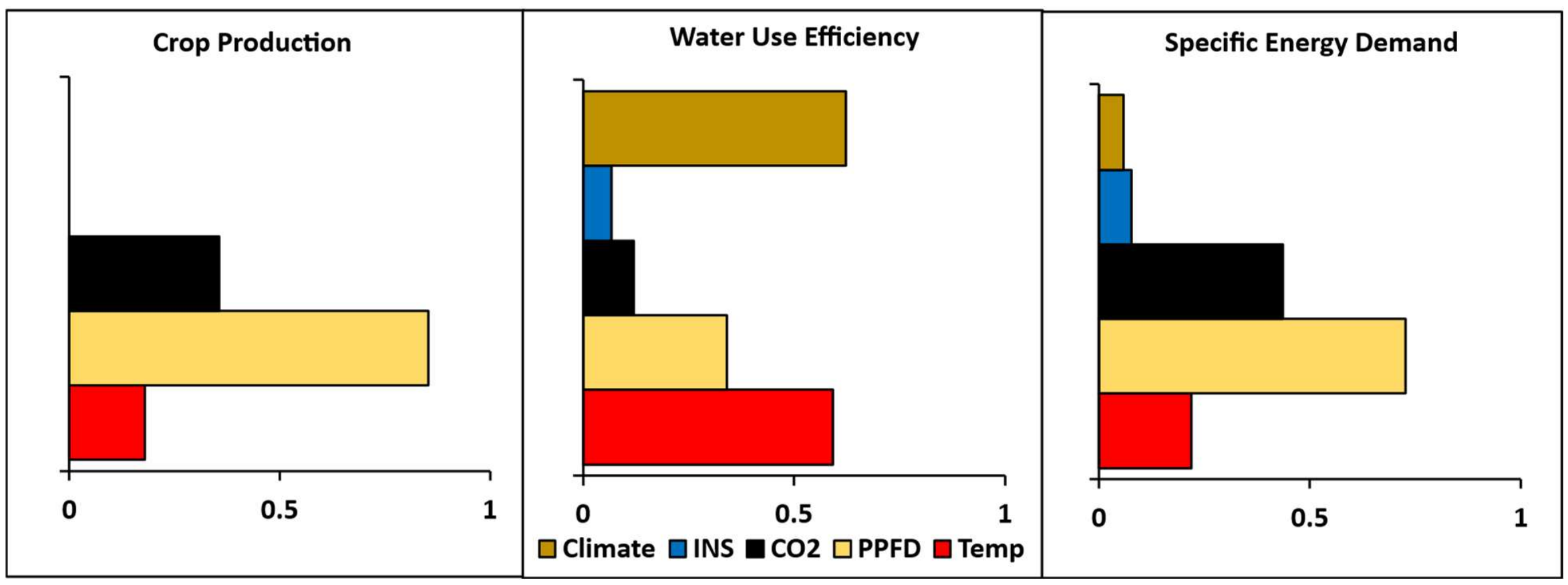

Figure 8: *Distance correlation coefficient of input parameters on resource consumption in vertical farming.*

### 3.5. Vertical Farm Electricity Consumption

The economic assessment of the simulated VFs required converting thermal energy demand into electrical load, achieved by modelling two air-to-water heat pumps (see Methods). One unit served heating and post-heating, while the other covered cooling and dehumidification; a single unit reversible unit was excluded due to the occasional simultaneity of heating and dehumidification. Using the developed algorithm, which accounts for COP variability with part-load factor, heat pump size, and external temperature, all thermal loads were converted into electrical loads, enabling estimation of the specific electrical energy consumption (SEEC) for each scenario.

Most of the considerations on the thermal loads also apply to electrical loads due to their close correlation. From a SEEC perspective, colder climates are more favourable: compared to Trondheim, SEC increased by 2.55% in Shanghai and 5.66% in Dubai, while SEEC rose by 7.8% and 8.14%, respectively. This is primarily because cooling load, the second largest energy demand after lighting, is more efficiently managed in colder climates where lower outdoor temperatures both reduce cooling requirements and improve heat pump performance.

The increase in COP due to colder climates is less pronounced than initially expected. For the same VF scenario, the cooling heat pump achieved COP values of 4.49 in Trondheim, 4.28 in Shanghai, and 3.56 in Dubai. The relatively small difference between Trondheim and Shanghai is attributed to the condenser temperature limitations implemented by manufacturers to avoid off-design compressor operation.

For these reasons, as shown in Figure *9*, the effect of the external climate on SEEC is relatively limited. Removing insulation generally increases SEEC, but this rise is partially mitigated by heat pump performance: heating COP is typically higher than cooling COP, and the heating heat pump often operates in part load conditions, further improving its COP. Consequently, the SEEC increase is limited to approximately +8% in Trondheim and +5.2% in Dubai. Due to negligible differences between Trondheim and Shanghai, only the former and Dubai are reported in Figure *9*. Similarly to the SEC, the most energy-efficient scenario is T_I_100_24_1400, with a SEEC of 5.05 $kWh_{el}\ kg^{-1}$.

Notably, this detailed analysis enabled a more accurate estimation of energy consumption compared to economic assessments based on a single average value. The model estimated a SEEC range of 5.05-15.81 $kWh_{el}$ $kg^{-1}$, whereas Moghimi et al. [17] assumed a fixed value of 5.75 kWh $kg^{-1}$, consistent, but limited. These findings highlight the impact of more robust models to enhance performance evaluations of vertical farming systems.

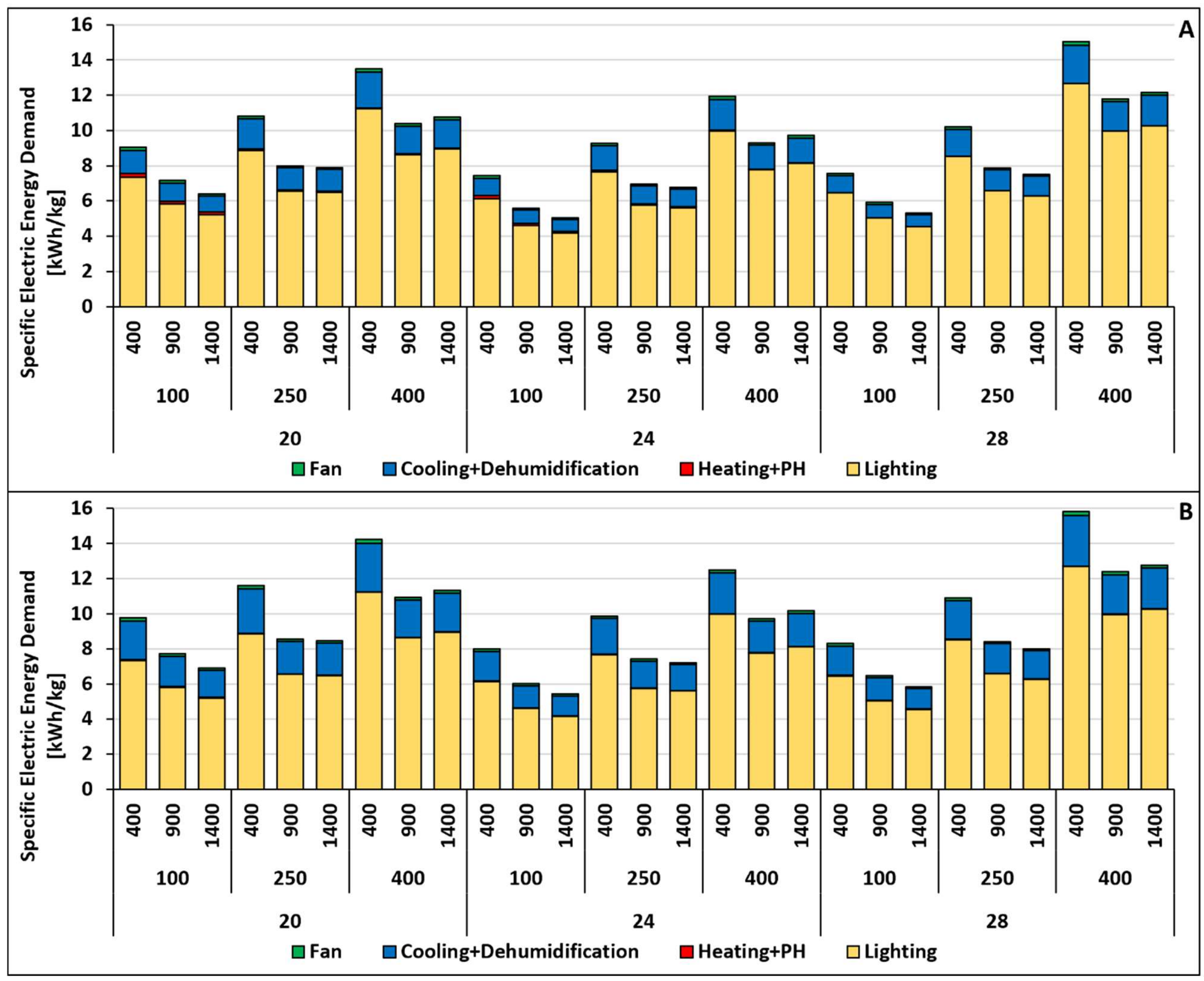

Figure 9: *Specific Electric Energy Consumption of the VF under two simulated locations: (A) Trondheim, (B) Dubai.*

### 3.6. Economic Assessments

Knowing the size of all the major components within the vertical farming system, the capital expenditure (CAPEX) for each scenario was calculated. Since the costs of all components are assumed to be independent of the VF's geographical location, only the CAPEX for the Trondheim site is shown in Figure *10*. The cost of the carbon fertilization system was considered negligible, so Figure *10* illustrates only the influence of indoor temperature and PPFD on the CAPEX.

Even from an economic perspective, it is noteworthy that the HVACD system contributes a relatively modest portion, representing approximately 14% of the total CAPEX. The cost of the growing chamber is fixed, as it depends solely on the net growing area, and accounts for 19 to 41% of the capital investment. It is important to note that the building is assumed to be pre-existing and used under a lease agreement; therefore, construction costs are not included.

As with the energy-related results, the lighting system represents the most significant share of the capital costs, ranging from 35 to 55%. This includes future replacement expenses, as LEDs are the component with the shortest operational lifespan. Considering all simulated scenarios, the total CAPEX for a vertical farm is estimated to range between 61.0 and 140.0 thousand dollars, resulting in a specific cost of approximately 680 to 1560 $ per square meter of cultivated area.

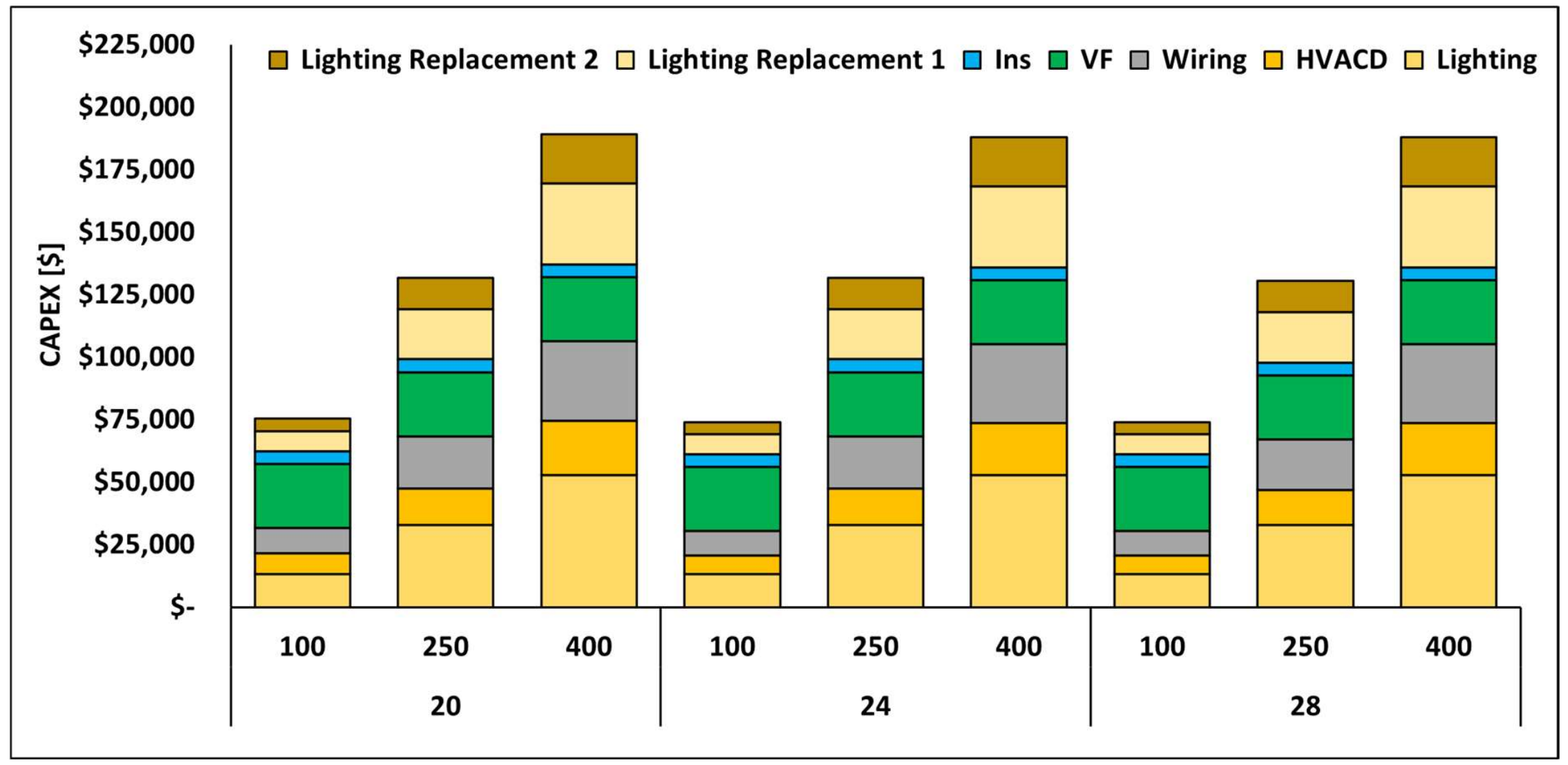

Figure 10: *Capital Expenditures of the well-insulated VF as a function of indoor conditions (Temperature and PPFD).*

The operational expenditure, which includes annual costs for water, $CO_2$, electricity, labour, and leasing, was evaluated for each considered location in order to highlight the differences (see Figure *11*). It is important to note that, in all scenarios, labour costs represent the largest portion of the annual expenses of the vertical farm, accounting for approximately 25% in Shanghai, 39% in Trondheim, and 35% in Dubai.

Due to variations in cost of living, such as leasing rates and salaries, Shanghai results as the most cost-efficient location among those analyzed in this study for implementing vertical farming, with an annual OPEX ranging from 11 to 34 k$. Under identical indoor conditions, the corresponding values for Trondheim and Dubai are 38 to 72% and 65 to 90% higher, respectively.

Because of the high energy demand required by vertical farming systems, the portion of OPEX related to electricity reaches about 12% when the light intensity is low and up to 19% when it is high, in the case of Trondheim. Although these values may vary slightly across locations, the general order of magnitude remains consistent.

Lastly, the contributions of carbon dioxide and water to the operational costs are minimal. The high water use efficiency achieved by the system significantly limits the need for external water supply, resulting in a nearly negligible cost for water.

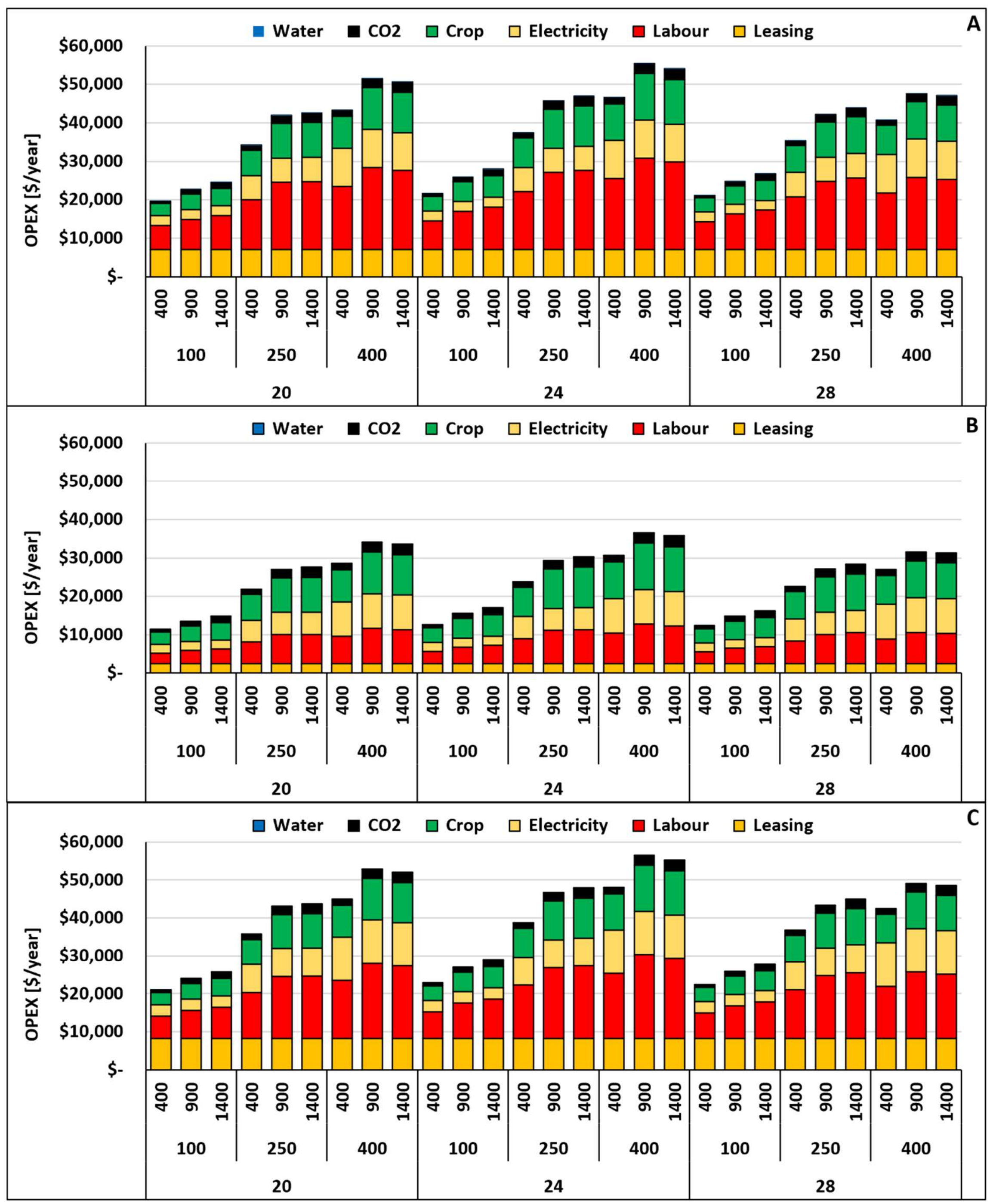

Figure 11: *Operational Expenditures of the well-insulated VF as a function of indoor conditions (Temperature, PPFD, and $CO_2$ concentration) in three simulated locations: (A) Trondheim, (B) Dubai.*

To present the overall results of the economic assessment, the levelized cost of lettuce for each insulated scenario is illustrated in Figure *12*. The trend of the LCoL remains consistent across all studied locations, as the productivity of the vertical farm is not influenced by external climatic conditions. As anticipated from the CAPEX and OPEX evaluations, the LCoL values for the vertical farm located in Shanghai are significantly lower than those observed in Trondheim and Dubai.

The $CO_2$ prices considered resulted in lower LCoL values in scenarios where higher $CO_2$ concentrations led to increased crop productivity. For example, at a PPFD of 250 µmol $m^{-2}$ $s^{-1}$, a 12% increase in $CO_2$ cost would make the 900 ppm and 1400 ppm concentrations equally cost-effective. However, in that same scenario, the $CO_2$ price would need to be approximately four and a half times higher to render carbon fertilization economically unviable. This threshold decreases at higher PPFD levels, while it increases at lower PPFD levels.

Instead, the electricity price directly influences the trade-off between the higher productivity and the higher energy consumption, both of which are associated with increasing the PPFD level. With the selected electricity price, the assessment showed that the lowest LCoL was achieved by setting an intermediate PPFD of 250 µmol $m^{-2}$ $s^{-1}$ across all studied locations. Specifically, in Trondheim, if the electricity price increased up to 0.51-1.85 \$ $kWh^{-1}$, the 100 µmol $m^2$ $s^1$ scenario would become competitive. Notably, none of the analyzed scenarios with a PPFD of 400 µmol $m^{-2}$ $s^{-1}$ proved cost-effective, even assuming free electricity for the VF. This is because the rise in CAPEX is not compensated for by the gain in crop productivity. Since light use efficiency and overall crop production showed opposite trends with PPFD, a trade-off is required to identify the most cost-competitive scenario. Table *5* summarizes this analysis for each indoor temperature and location evaluated, assuming a $CO_2$ concentration of 900 ppm. The terms $C_{el400}$ and $C_{el100}$ refer to the electricity prices required to make the high and low PPFD scenarios, respectively, as viable as the intermediate one. However, electricity price is not the only factor: crop-related OPEX, such as seed, nutrients, and labour, also influence the optimal scenario. Increased automation could lower labour costs, enabling low electricity prices to make higher-PPFD scenarios more competitive.

Based on the presented results, the most cost-effective scenarios among those analyzed in this study are T_I_250_24_1400, S_I_250_24_1400, and D_I_ 250_24_1400, achieving a LCoL of 6.38, 4.57, and 6.48 \$ $kg^{-1}$, respectively. It is noteworthy that, with an installation cost of 5.1 k\$, the insulation slightly increases the LCoL when compared with the corresponding uninsulated scenario. In a cold climate such as Trondheim, if the insulation cost rises by more than 35%, the uninsulated scenario becomes more cost-effective. A similar analysis in a hot climate, such as Dubai, shows that a cost increase of only 20% would make the insulated configuration less economically favourable. This difference is influenced not only by the external climate but also by other location-dependent expenses, including leasing, labour, and electricity.

The electricity component of the LCoL ranges from 0.46 to 1.74 \$ $kg^{-1}$, consistent with the values reported by Arcasi et al. [18], considering differences in LED efficiency assumptions. However, the higher overall LCoL values emphasize the need for comprehensive economic analyses to assess the profitability of vertically farmed lettuce. Moghimi et al. [17] reported an average LCoL of 1.48 \$ $kg^{-1}$, considerably lower than the values estimated in this study. That value was based on energy and labor costs of 0.46 \$ $kg^{-1}$ (within the range) and 1 \$ $kg^{-1}$, respectively. In this study, labor costs ranged from 0.97 to 2.23 $\$\ kg^{-1}$, underscoring the influence that CO2, CAPEX, and leasing expenditures on the final price of the lettuce and highlighting the contribution of this work.

Table 5: *Electricity price variations required to improve the competitiveness of high-PPFD scenarios.*

| **Location** | **Temperature [°C]** | $C_{el400}\ [\$\ kWh^{-1}]$ | $C_{el100}\ [\$\ kWh^{-1}]$ |
|---|---|---|---|
| | 20 | -0.04 | 1.85 |
| Trondheim | 24 | -0.06 | 0.73 |
| | 28 | -0.13 | 0.51 |
| | 20 | -0.08 | 1.00 |
| Shanghai | 24 | -0.09 | 0.36 |
| | 28 | -0.13 | 0.24 |
| | 20 | -0.03 | 2.14 |
| Dubai | 24 | -0.05 | 0.82 |
| | 28 | -0.12 | 0.58 |

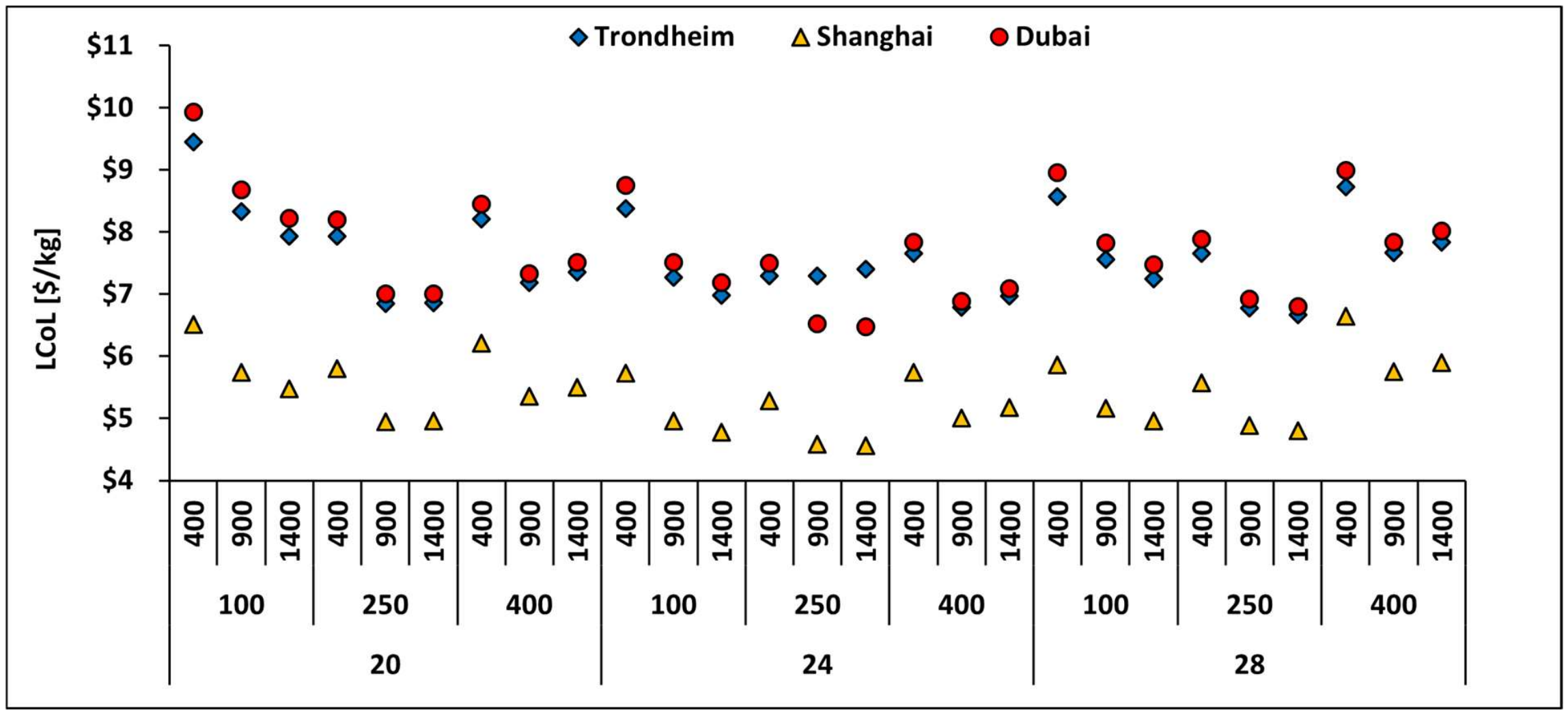

Figure 12: *Levelized cost of lettuce for a well-insulated VF under varying indoor conditions and external socio-environmental settings.*

### 3.7. Sustainability Assessment

After evaluating the vertical farm's performance from agricultural, energy, and economic perspectives, it is important to assess its sustainability by comparing vertically grown lettuce with imported lettuce. Since the previous section identified the most cost-effective scenarios, these have been selected for comparison in terms of $CO_2$ equivalent emissions with those associated with an equivalent quantity of imported lettuce.

In this analysis, only the emissions associated with the energy demand of the vertical farm have been considered. Figure *13* illustrates the carbon footprint savings of vertically grown lettuce compared to imported lettuce across the three studied locations. These savings are primarily influenced by the local energy mix and the characteristics of the supply chain. Under current conditions, it becomes evident that highly energy-intensive vertical farms can only be considered environmentally sustainable in regions where the share of renewable energy in the grid is substantial. In Norway, where hydropower and wind energy account for 89% and 9% of the energy mix, respectively, vertical farming achieves a carbon reduction of 230 grams of $CO_2$ per kilogram of lettuce. This corresponds to a 70% decrease in emissions compared to the import alternative.

In Shanghai and Dubai, the high emission factors associated with electricity consumption make vertical farming significantly unsustainable from a carbon footprint perspective. Adopting solutions with lower light intensity would improve the sustainability of vertical farming in these contexts by reducing emissions by 60 percent when switching the PPFD from 250 to 100 µmol $m^{-2}$ $s^{-1}$. Due to China's high emission factor and the potentially short supply chain, an improvement in energy efficiency of 98.8% would be necessary to make vertical farming sustainable. In Dubai, where the supply chain is more emission-intensive, this required efficiency improvement is reduced to 62%.

The results show how vertical farms can contribute to sustainable agriculture and urban food systems by combining productivity gains with lower resource use. The strong role of lighting and $CO_2$ enrichment highlights the importance of technological innovation, while the influence of local energy and economic contexts links vertical farming to broader goals of the energy transition and resilient urban supply chains.

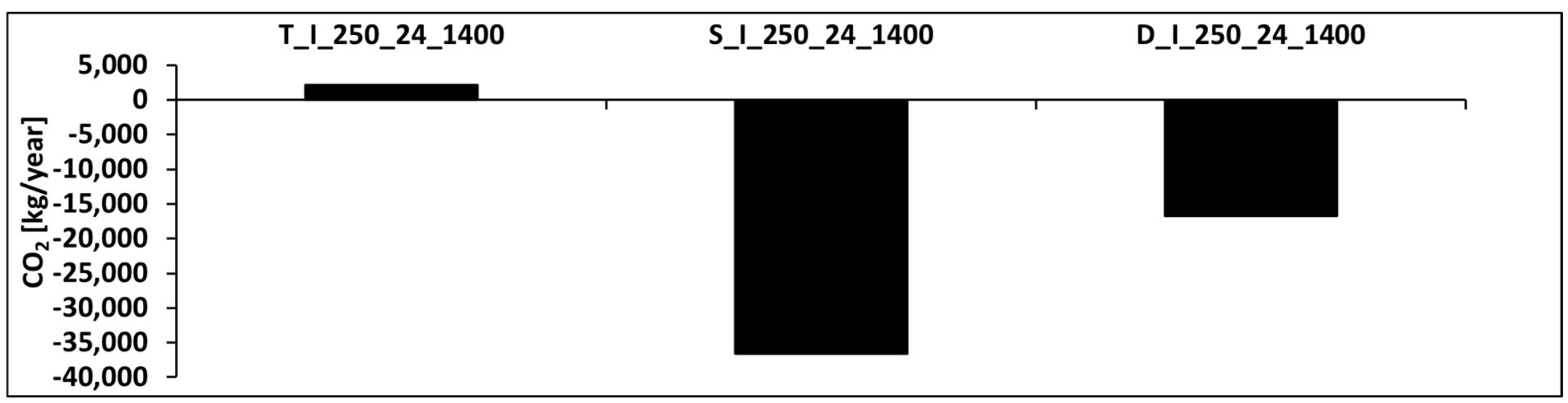

Figure 13: *Estimated $CO_2$-eq emission savings from local vertical farm (VF) lettuce production versus conventional supply chain alternatives.*

### 3.8. Distance Correlation for the Economic Assessments

This section assesses the influence of input variables on the investment requirements (CAPEX), operational costs, LCoL, and $CO_2$-eq emission savings (Figure *14*). Overall, indoor temperature and insulation have negligible effects on costs and emissions, reflecting their minor role in resource consumption and crop productivity.

PPFD strongly affects CAPEX ($\rho$ = 0.999) and OPEX ($\rho$ = 0.730) since light intensity drives both system size and energy use. However, its impact on the LCoL is modest ($\rho$ = 0.200), as productivity partly offsets higher costs. Conversely, $CO_2$ fertilization directly enhances crop growth, improving both LCoL ($\rho$ = 0.327) and equivalent $CO_2$ emission savings ($\rho$ = 0.448), while slightly impacting OPEX ($\rho$ = 0.222). Climate, understood as both environmental and socio-economic conditions, exerts limited influence on CAPEX but plays a major role in OPEX ($\rho$ = 0.496), LCoL ($\rho$ = 0.734), and $CO_2$-eq savings ($\rho$ = 0.576). This is mainly due to differences in electricity prices, labour costs, land costs, and supply across regions.

From a practical standpoint, these results suggest that entrepreneurs should carefully balance lighting intensity against productivity gains, as higher CAPEX and OPEX do not always translate into better unit economics. At the same time, $CO_2$ enrichment appears to be a cost-effective strategy to lower production costs and carbon intensity. For policymakers, the strong effect of the local socio-economic context highlights the importance of energy pricing policies, labor market conditions, and grid decarbonization strategies in shaping the competitiveness and sustainability of vertical farms.

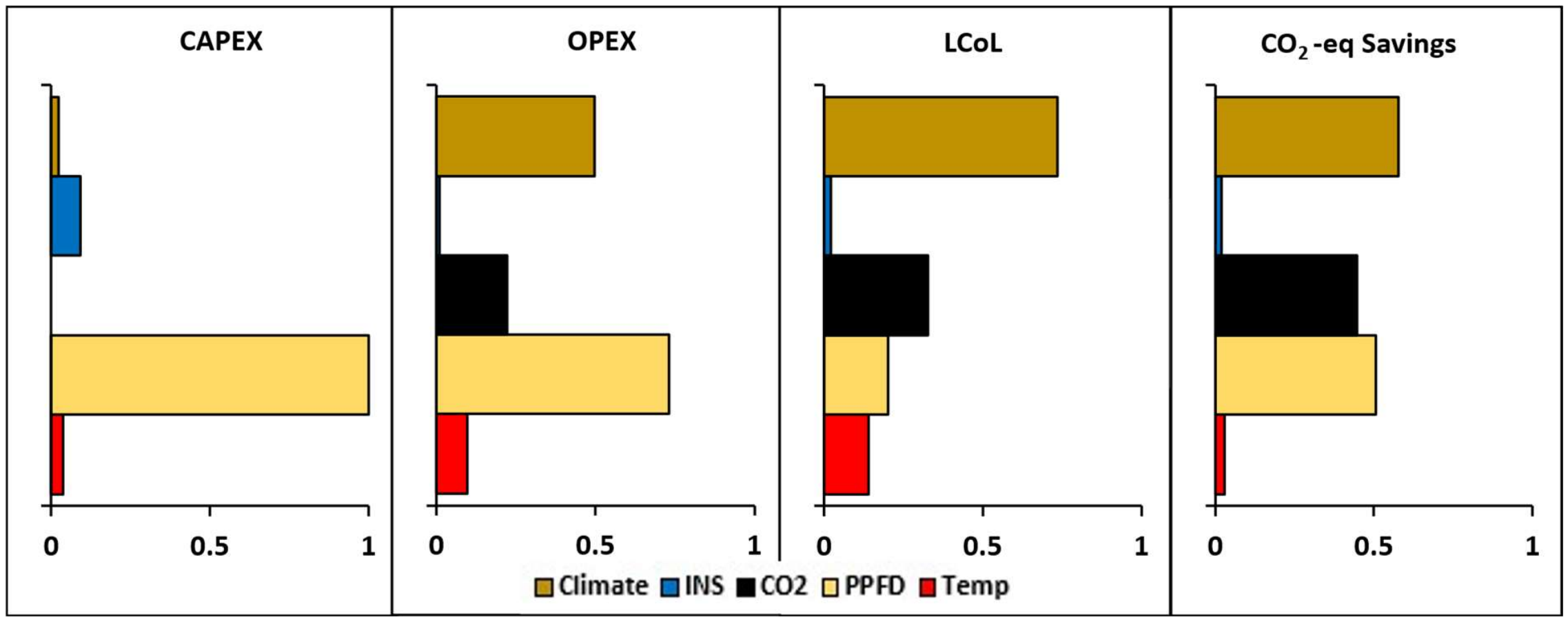

Figure 14: *Distance correlation coefficient of input parameters on economic and sustainability performance in vertical farming.*

## 4. Conclusions

This study provides a novel, data-driven framework for designing hydroponic vertical farming by integrating crop modeling, energy use, economic assessment, and environmental impact. By capturing complex non-linear relationships between input parameters and system performance, it identifies context-specific strategies to maximize productivity while minimizing energy costs and carbon footprint. All the annual resource requirements obtained by the employed transient model were subsequently integrated into the economic assessment to estimate both capital and operational costs, leading to the calculation of the levelized cost of lettuce for each scenario analyzed. Finally, different supply chain conditions were considered to compare the emissions associated with vertical farming production against those from importing an equivalent amount of lettuce.

To assess the influence of indoor conditions, insulation, and VF's location on key performance indicators, 162 one-year scenarios were run varying temperature, PPFD, $CO_2$ concentration, insulation, and climate (Trondheim, Shanghai, Dubai). Distance correlation coefficients were used to quantify each parameter's impact.

The main conclusions of this analysis are summarised as follows:

- The HVACD system maintained stable productivity across climates. Crop growth was mainly driven by PPFD ($\rho$ = 0.85), with $CO_2$ ($\rho$ = 0.36) and temperature ($\rho$ = 0.22). Productivity peaked at 400 µmol $m^{-2}$ $s^{-1}$, but lower LUE at high PPFD highlights the need to balance yield and energy costs. $CO_2$ fertilization saturated at 900 ppm, and 24°C was identified as the optimal temperature for lettuce.
- Considering the trade-off between energy use and productivity, the LCoL was the key metric for identifying optimal conditions. The lowest LCoL (4.57 to 6.48 \$ $kg^{-1}$ across locations) occurred at 24°C, PPFD of 250 μmol $m^{-2}$ $s^{-1}$, $CO_2$ 1400 ppm, and with insulation.

- In cold climates, removing insulation reduced the SEC by up to 5% under high PPFD, as heat losses offset cooling demand; however, at low PPFD, heating needs raised the SEC by 40-70%. In hot climates, the absence of insulation consistently increased SEC (1-21%), under high PPFD and elevated temperatures, partly mitigating these effects.
- Electricity price moderately influenced LCoL (ρ = 0.22), but high-PPFD scenarios remained economically unfeasible even with free electricity, as CAPEX increases outweigh productivity gains. Greater automation could shift this balance by lowering labour costs. Insulation had a minor impact (1 to 3 c$ $kg^{-1}$), yet its adoption becomes unjustified if installation costs exceed reference values by 20-35%.
- Finally, its carbon footprint is highly sensitive to the local energy mix (ρ = 0.58). Currently, only countries with near-complete energy decarbonization, such as Norway, can support vertical farming without increasing $CO_2$ emissions compared to conventional lettuce imports.
- Notably, in hot and arid climates, vertical farms achieved higher WUE due to the moisture condensation from the incoming air within the HVACD system. This process may significantly reduce water demand in regions where freshwater is scarce and would otherwise require energy-intensive desalination for greenhouse farming.

These findings can directly support policymakers in evaluating the environmental feasibility of vertical farming under different energy mixes and climate conditions, and guide practitioners in selecting optimal indoor conditions, lighting strategies, $CO_2$ levels, and insulation measures to achieve cost-effective and sustainable production. A further contribution of this study lies in enriching the existing literature by providing more robust and reliable results through the use of a detailed transient model combined with distance correlation analysis. Future research should extend this approach to other crop types and explore the potential of vertical farming in humid climates, where reduced freshwater demand could further lower overall emissions.

## 5. Appendix and supplementary data

### 5.1 Distance Correlation Coefficient Method

The distance correlation coefficient (ρ) is a statistical measure used to evaluate the dependence between two different variables, X and Y, by comparing the similarities between their respective distance matrices, A and B. Let xi,j and yi,j represent the values taken by variables X and Y. The generic distance matrices Dij are defined as shown in Eqs. A1 and A2. Both matrices have dimensions n x n, where n refers to the number of observations, which in this study corresponds to the number of evaluated scenarios (162).

$$D_{ij}^{X} = |x_i - x_j| \tag{A1}$$

$$D_{ij}^{Y} = |y_i - y_j| \tag{A2}$$

Each distance matrix has been centered using a double centering operation, as described in Eqs. A3 and A4. In this process, $\overline{D}_{i.}, \overline{D}_{.j}, \overline{D}_{..}$ represent the average values calculated across the i-th row, the j-th column, and the entire matrix, respectively.

$$A_{ij} = \overline{D}_{ij}^{X} - \overline{D}_{i.}^{X} - \overline{D}_{.j}^{X} + \overline{D}_{..}^{X} \tag{A3}$$

$$B_{ij} = \bar{D}_{ij}^{Y} - \bar{D}_{i.}^{Y} - \bar{D}_{.j}^{Y} + \bar{D}_{..}^{Y} \quad \text{(A4)}$$

These matrices have ultimately been used to determine the covariance between X and Y, as well as the individual variances of the two selected variables, as expressed in Eqs. A5, A6, and A7. These expressions were employed in the equation reported in the methodology section.

$$dCov^2(X,Y) = \frac{1}{n^2}\sum_{i=1}^{n}\sum_{j=1}^{n} A_{ij}B_{ij} \quad \text{(A5)}$$

$$dVar^2(X) = \frac{1}{n^2}\sum_{i=1}^{n}\sum_{j=1}^{n} A_{ij}A_{ij} \quad \text{(A6)}$$

$$dVar^2(Y) = \frac{1}{n^2}\sum_{i=1}^{n}\sum_{j=1}^{n} B_{ij}B_{ij} \quad \text{(A6)}$$

## 5.2 Additional Data

To clarify the assumptions regarding the impact of insulation on vertical farm energy consumption, the wall stratification and related properties are summarized in Table A1.

*Table A1: Wall's stratification and related properties.*

| Layers | Plaster | Bricks | PUR | Plaster |
|---|---|---|---|---|
| **Thickness [mm]** | 20 | 250 | 105 | 10 |
| **Thermal Conductivity [W $m^{-1}$ $K^{-1}$]** | 0.870 | 0.720 | 0.022 | 0.520 |
| **Density [kg $m^{-3}$]** | 1800 | 1800 | 40 | 1400 |
| **Thermal Capacity [J $kg^{-1}$ $K^{-1}$]** | 840 | 1000 | 1450 | 1000 |

To illustrate the climatic differences among the three study locations, key data such as external temperature and global radiation are reported in Figure A1.

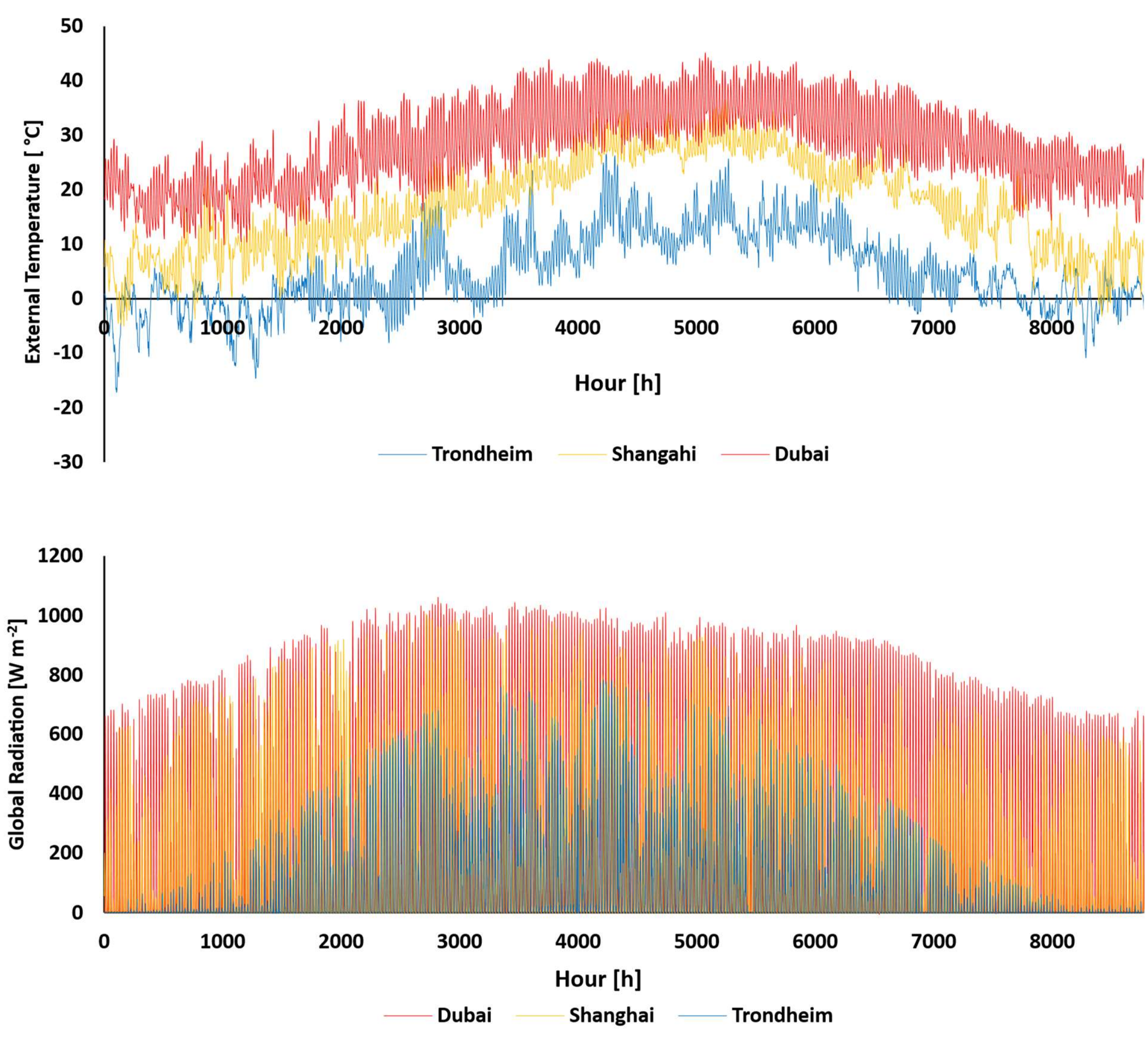

Figure A1: Climatic data across the simulated locations [28].

### 5.3 Relative Weight Analysis

This section presents a weight analysis of individual costs on CAPEX and OPEX to assess how relative uncertainties affect the outputs. Table A2 and Table A3 show the range of each expense's contribution across all simulated scenarios.

*Table A2: Impact of uncertainties related to individual cost on CAPEX across different simulated PPFD.*

| PPFD [µmol $m^{-2}$ $s^{-1}$] | Lighting System | HVACD | Wiring | Growing Chamber | Insulation |
|---|---|---|---|---|---|
| **100** | 21% | 13% | 16% | 41% | 8% |
| **250** | 33% | 15% | 33% | 26% | 5% |
| **400** | 39% | 16% | 51% | 19% | 4% |

*Table A3: Impact of uncertainties related to individual cost on OPEX across different locations simulated.*

| Location | $CO_2$ | Crop | Electricity | Water | Labor | Leasing |
|---|---|---|---|---|---|---|

| **Trondheim** | 3-7% | 16-22% | 9-25% | 0% | 31-44% | 14-36% |
|---|---|---|---|---|---|---|
| **Shanghai** | 5-11% | 28-35% | 16-29% | 0% | 24-30% | 7-22% |
| **Dubai** | 3-6% | 15-22% | 12-27% | 0% | 28-40% | 15-39% |

### 5.3 Thermal Load of Insulated Vertical Farms

This section presents the thermal load trends for all simulated scenarios, enabling comparison between them and highlighting the impact of climate, PPFD, and temperature on the thermal load when the building is well insulated. It is noteworthy that in all scenarios, the heating load decreases as the external climate becomes warmer. Additionally, as PPFD increases, the cooling load rises significantly, while the dehumidification load becomes nearly negligible.

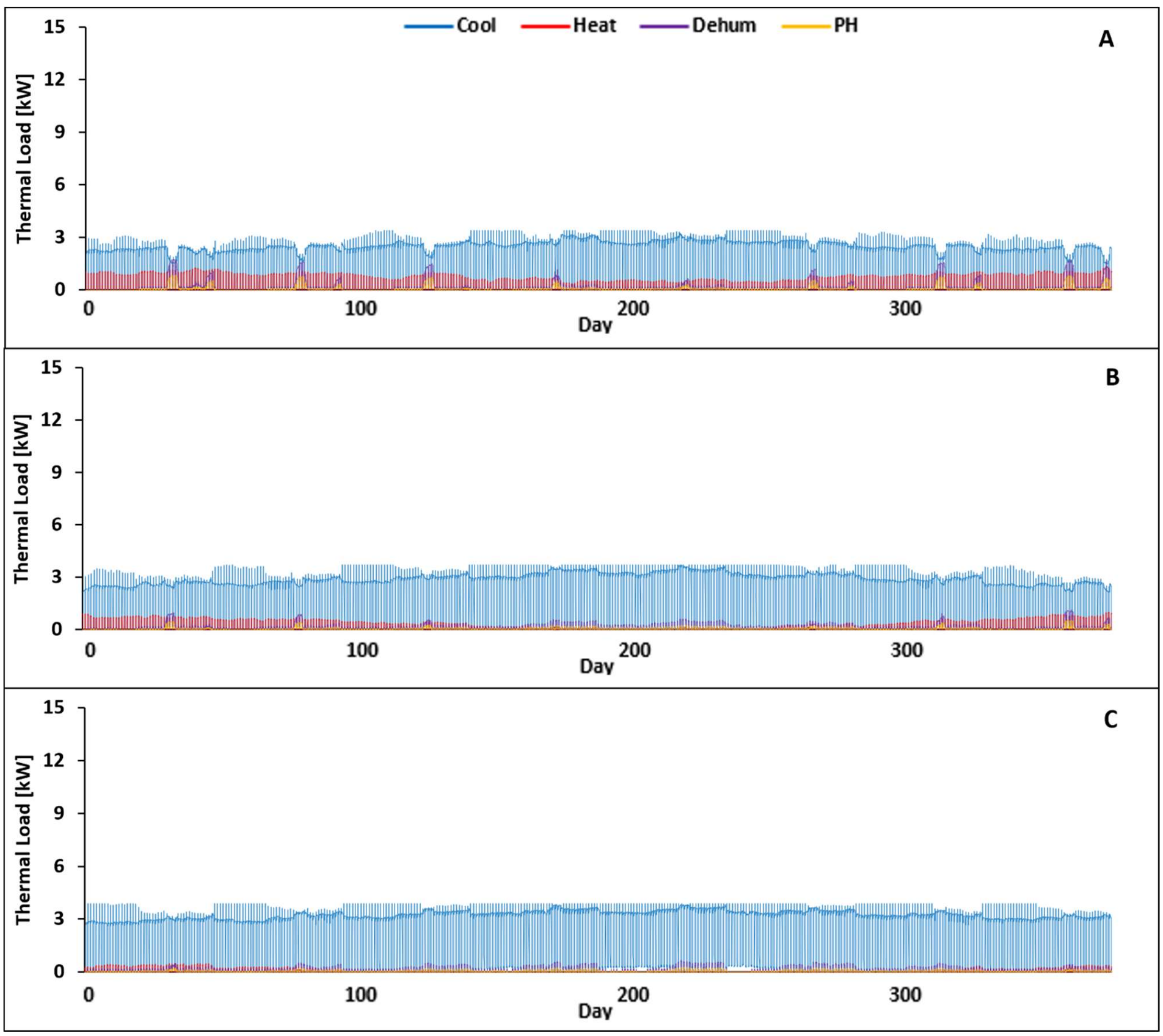

Figure A2: *Heating, cooling, dehumidification and post-heating loads throughout the year for a well-insulated VF in three locations: (A) Trondheim, (B) Shanghai, (C) Dubai. The results were obtained under conditions of 24 °C and 100 µmol $m^{-2}$ $s^{-1}$.*

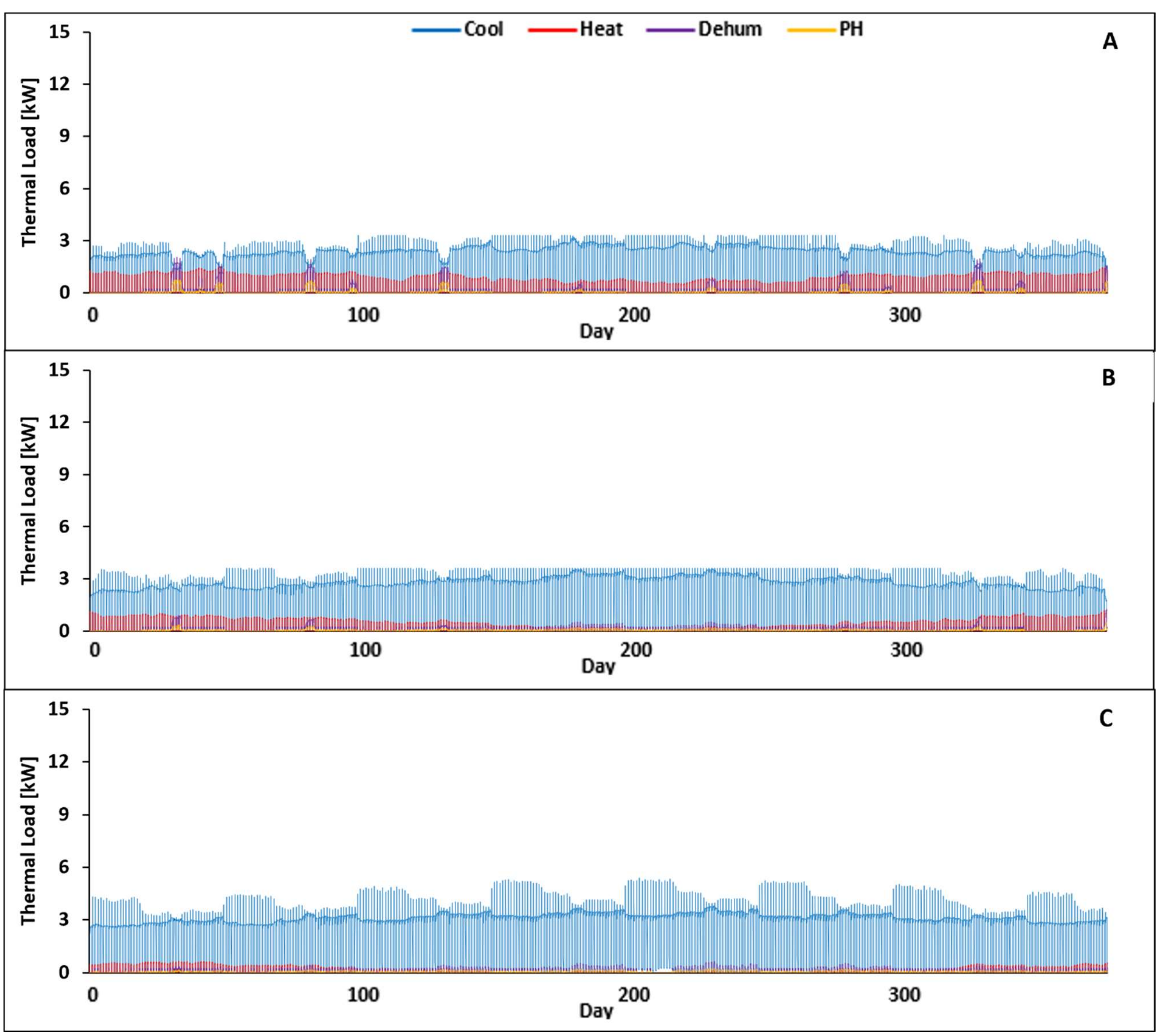

*Figure A3: Heating, cooling, dehumidification and post-heating loads throughout the year for a well-insulated VF in three locations: (A) Trondheim, (B) Shanghai, (C) Dubai. The results were obtained under conditions of 28 °C and 100 μmol m$^{-2}$ s$^{-1}$.*

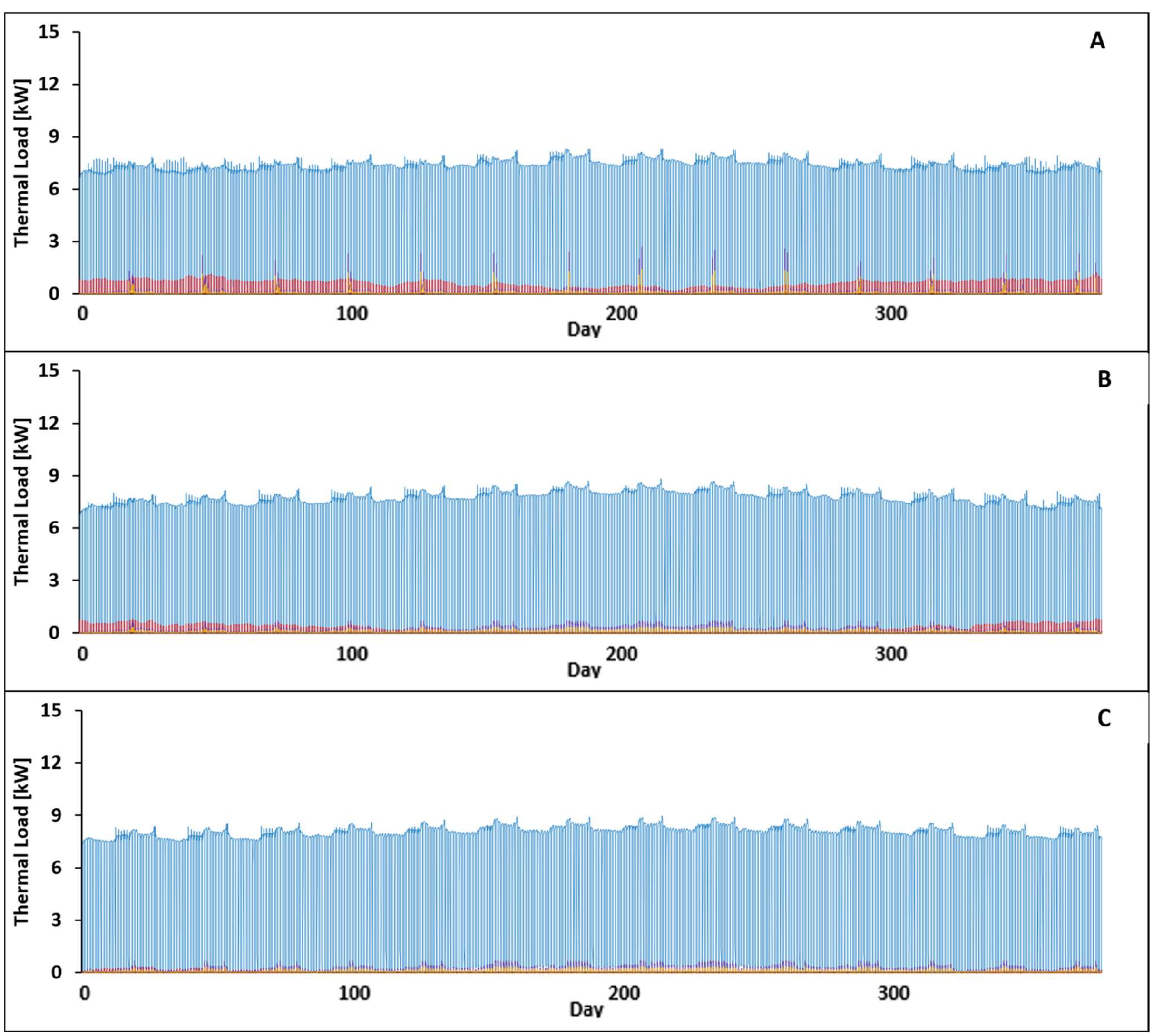

*Figure A4: Heating, cooling, dehumidification and post-heating loads throughout the year for a well-insulated VF in three locations: (A) Trondheim, (B) Shanghai, (C) Dubai. The results were obtained under conditions of 20 °C and 250 µmol $m^{-2}$ $s^{-1}$.*

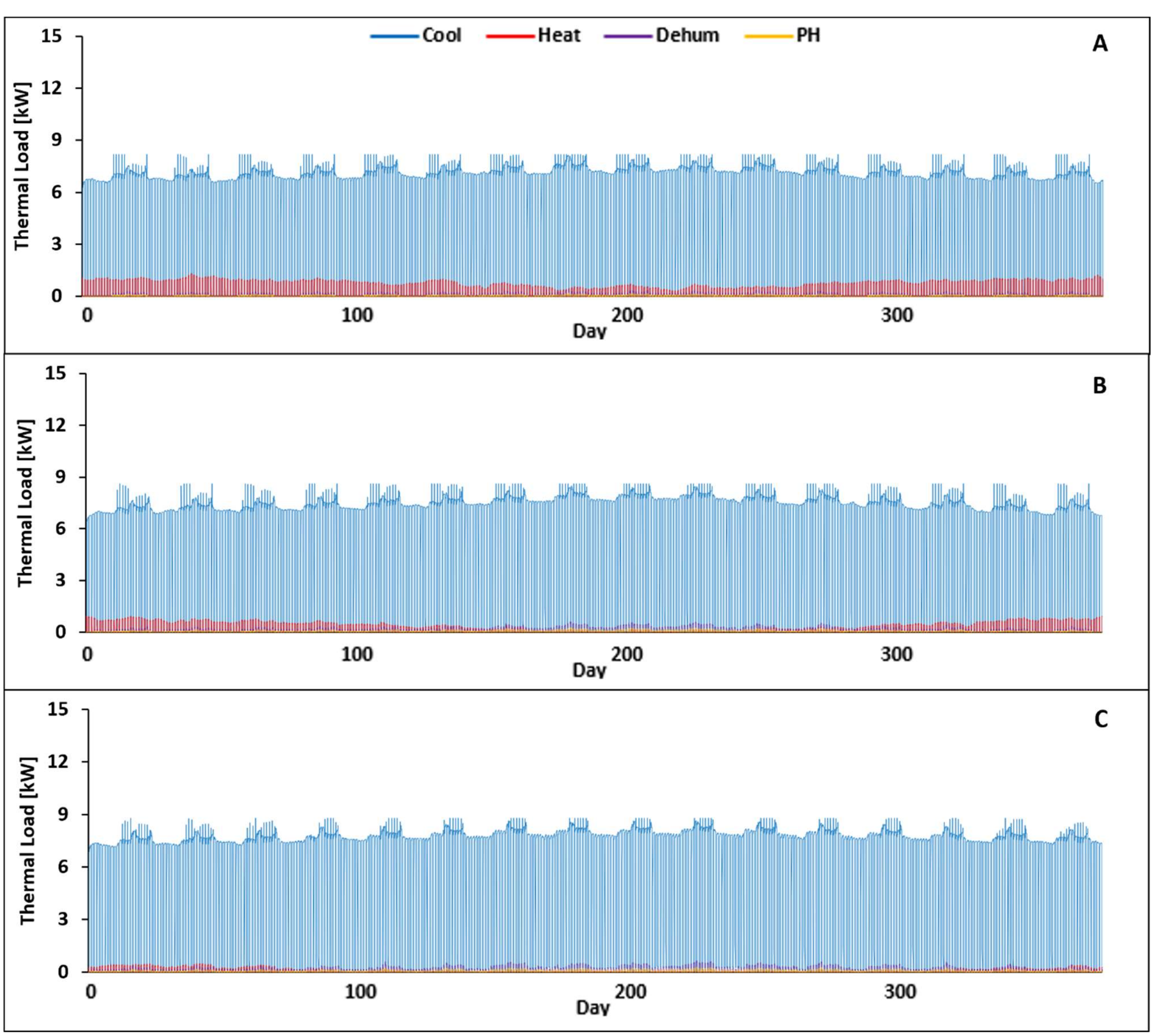

*Figure A5: Heating, cooling, dehumidification and post-heating loads throughout the year for a well-insulated VF in three locations: (A) Trondheim, (B) Shanghai, (C) Dubai. The results were obtained under conditions of 24 °C and 250 µmol m<sup>-2</sup> s<sup>-1</sup>.*

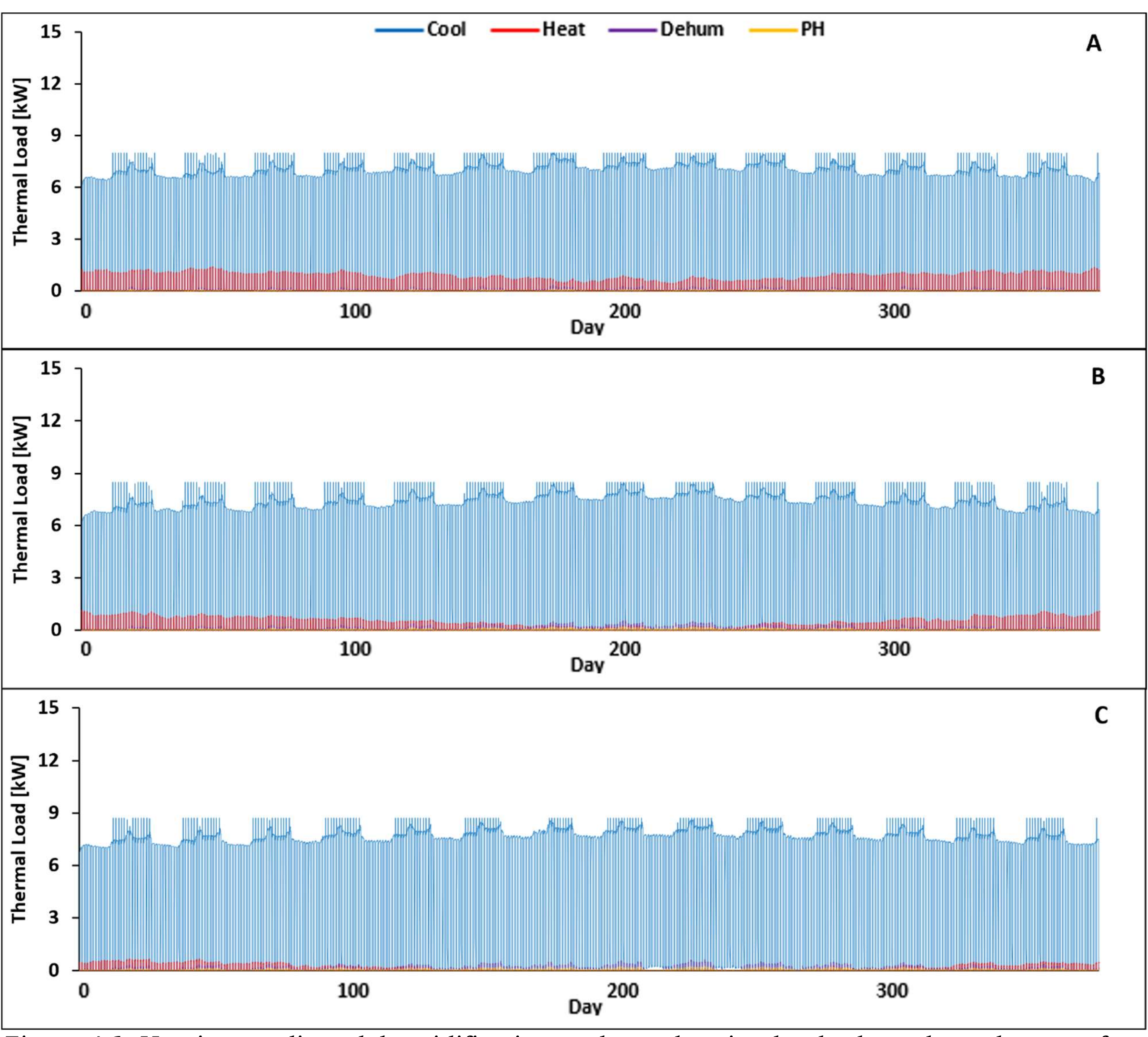

*Figure A6: Heating, cooling, dehumidification and post-heating loads throughout the year for a well-insulated VF in three locations: (A) Trondheim, (B) Shanghai, (C) Dubai. The results were obtained under conditions of 28 °C and 250 µmol m$^{-2}$ s$^{-1}$.*

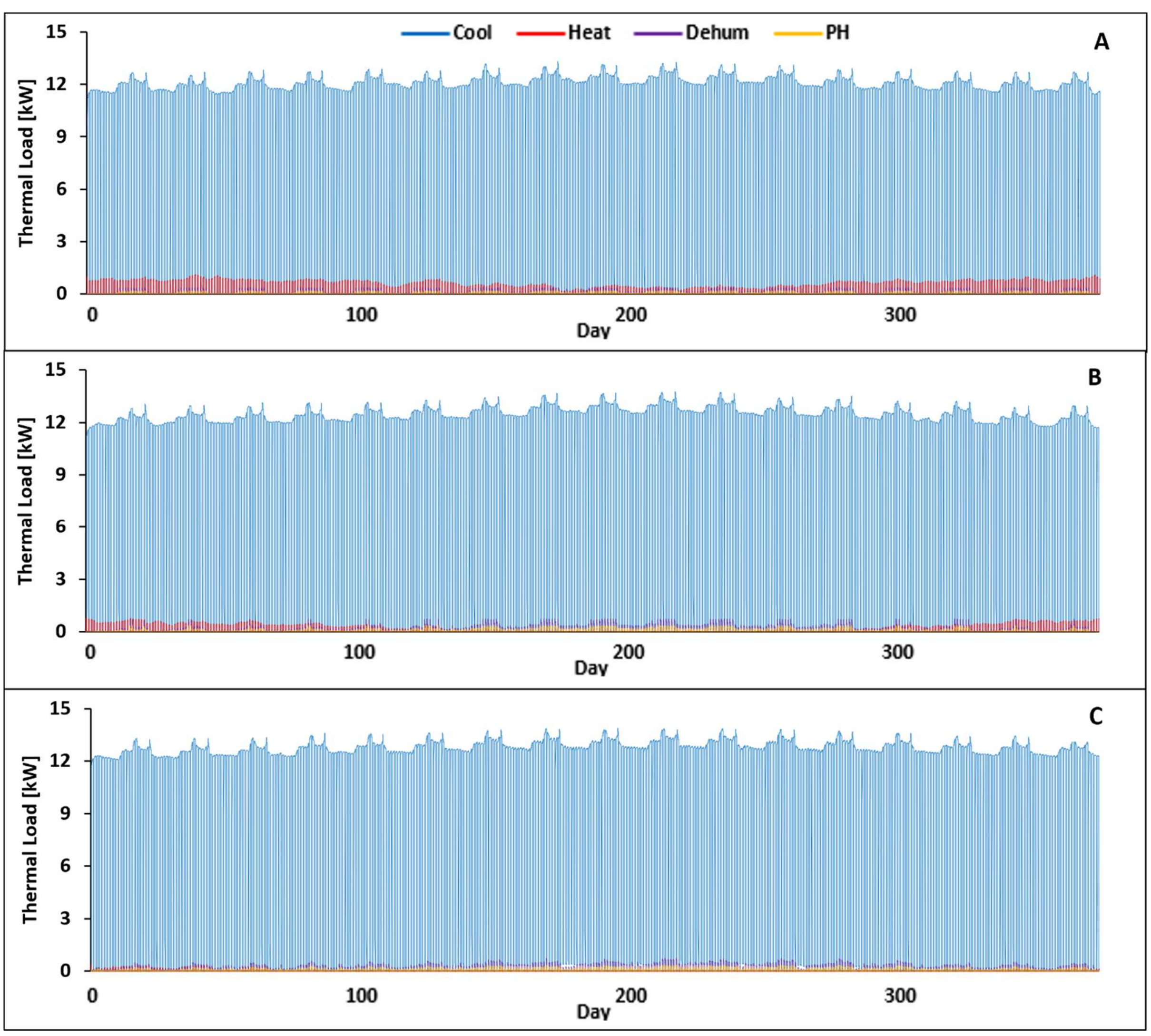

*Figure A7: Heating, cooling, dehumidification and post-heating loads throughout the year for a well-insulated VF in three locations: (A) Trondheim, (B) Shanghai, (C) Dubai. The results were obtained under conditions of 20 °C and 400 µmol m$^{-2}$ s$^{-1}$.*

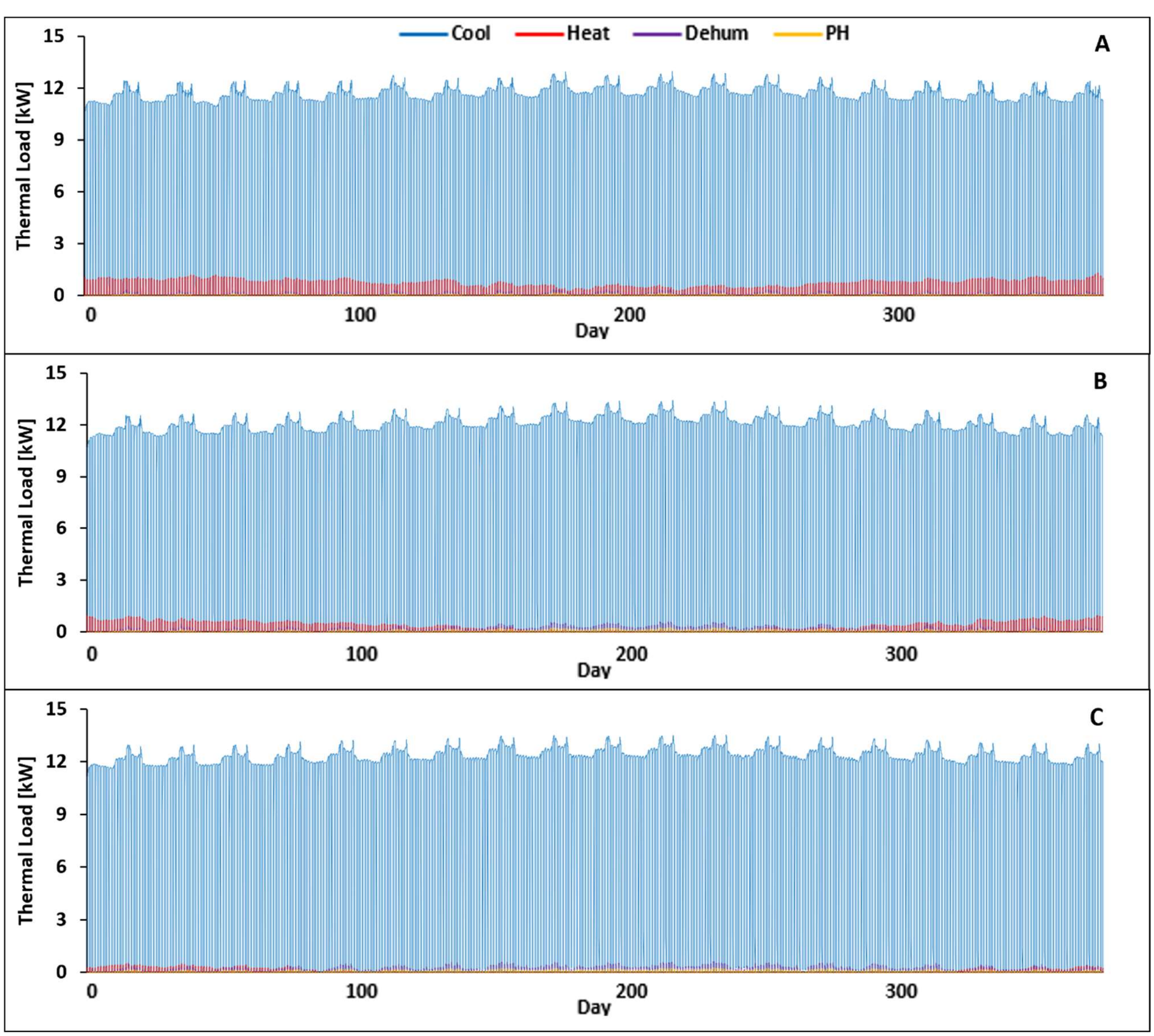

*Figure A8: Heating, cooling, dehumidification and post-heating loads throughout the year for a well-insulated VF in three locations: (A) Trondheim, (B) Shanghai, (C) Dubai. The results were obtained under conditions of 24 °C and 400 μmol $m^{-2}$ $s^{-1}$.*

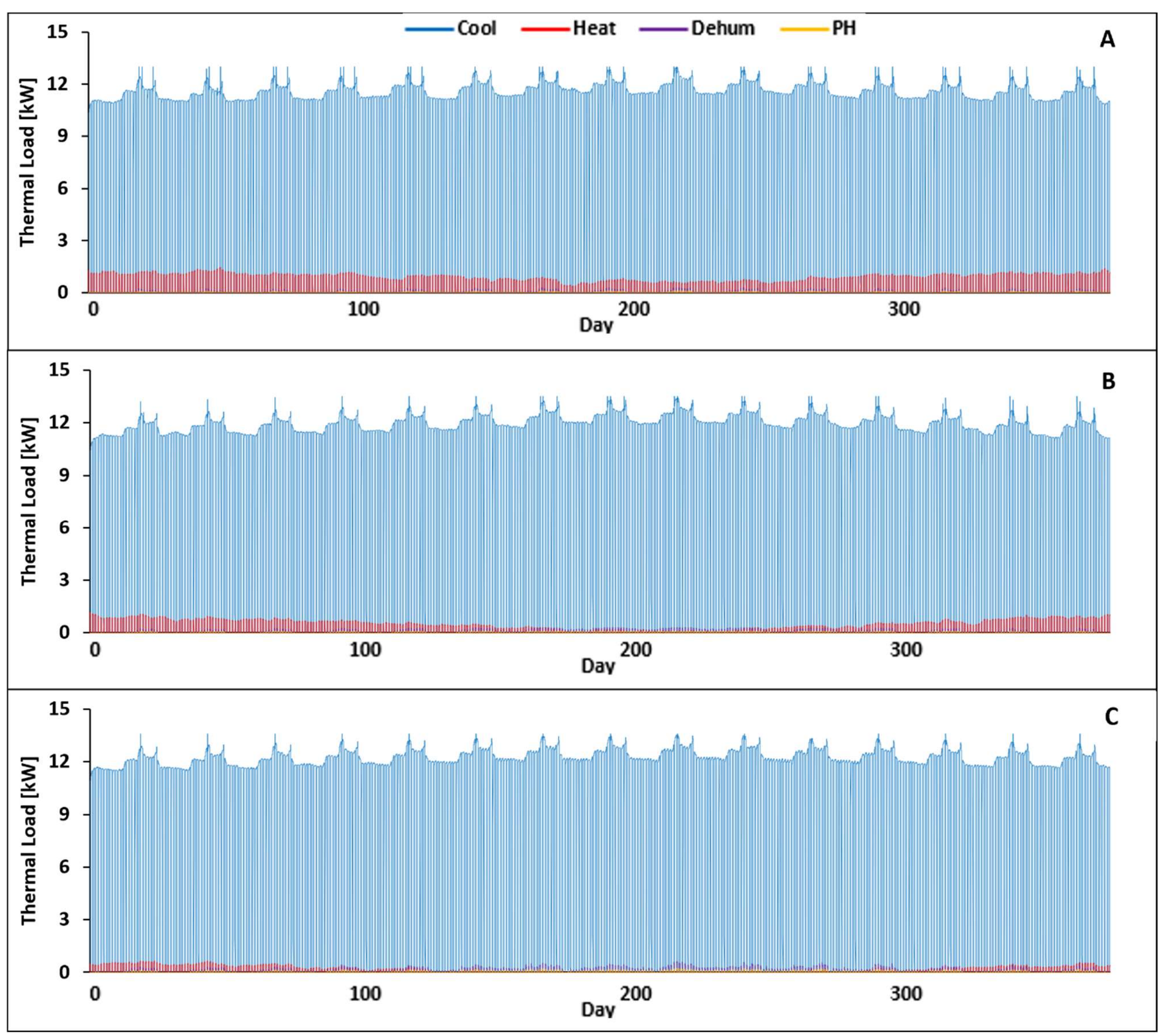

*Figure A9: Heating, cooling, dehumidification and post-heating loads throughout the year for a well-insulated VF in three locations: (A) Trondheim, (B) Shanghai, (C) Dubai. The results were obtained under conditions of 28 °C and 400 µmol m$^{-2}$ s$^{-1}$.*

## 5.4 Thermal Load of Not-Insulated Vertical Farm

This section presents the thermal load trends for all simulated scenarios without the insulation. The observations made in the previous section also apply to these non-insulated scenarios.

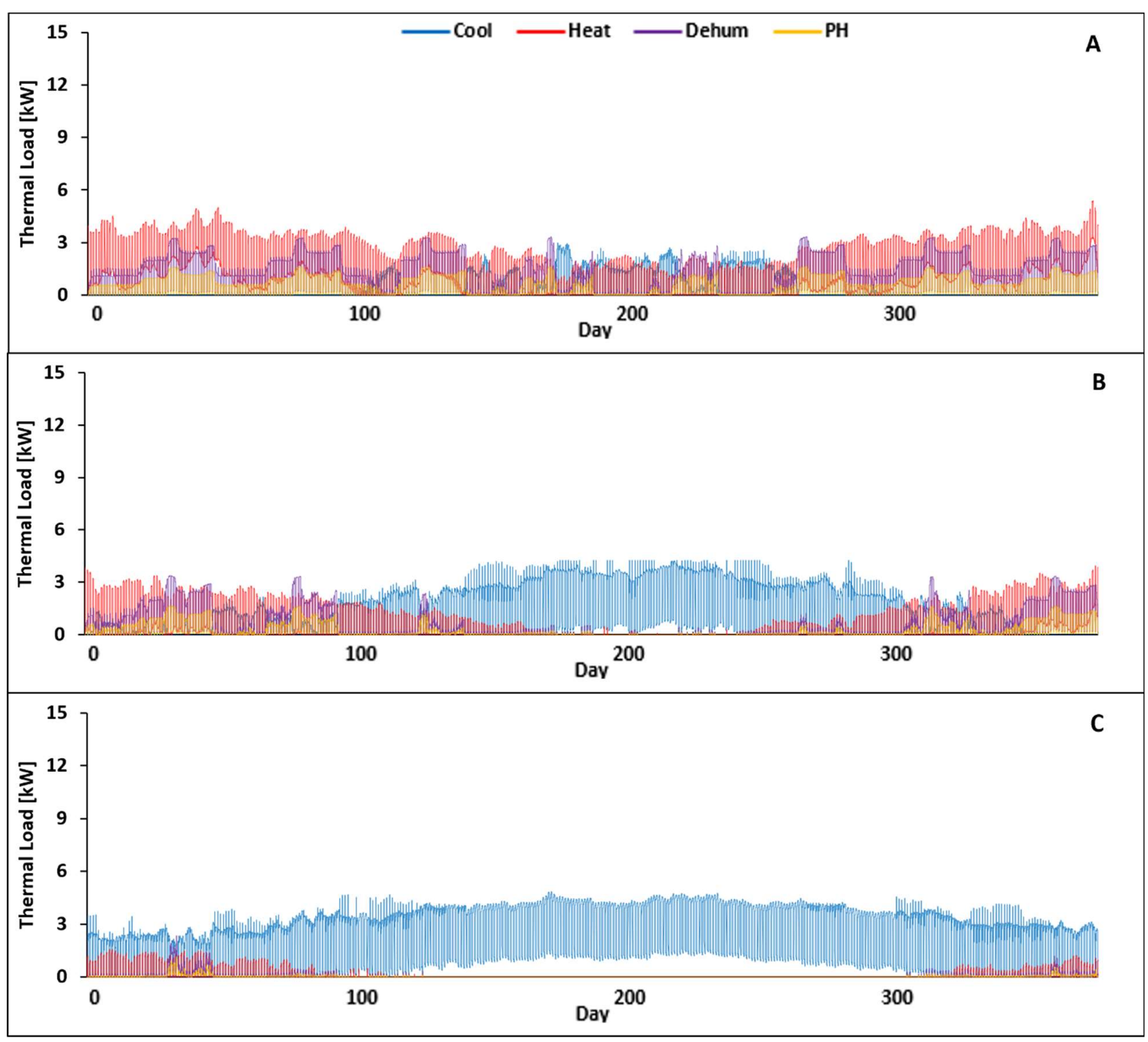

*Figure A10: Heating, cooling, dehumidification and post-heating loads throughout the year for a not-insulated VF in three locations: (A) Trondheim, (B) Shanghai, (C) Dubai. The results were obtained under conditions of 24 °C and 100 µmol m$^{-2}$ s$^{-1}$.*

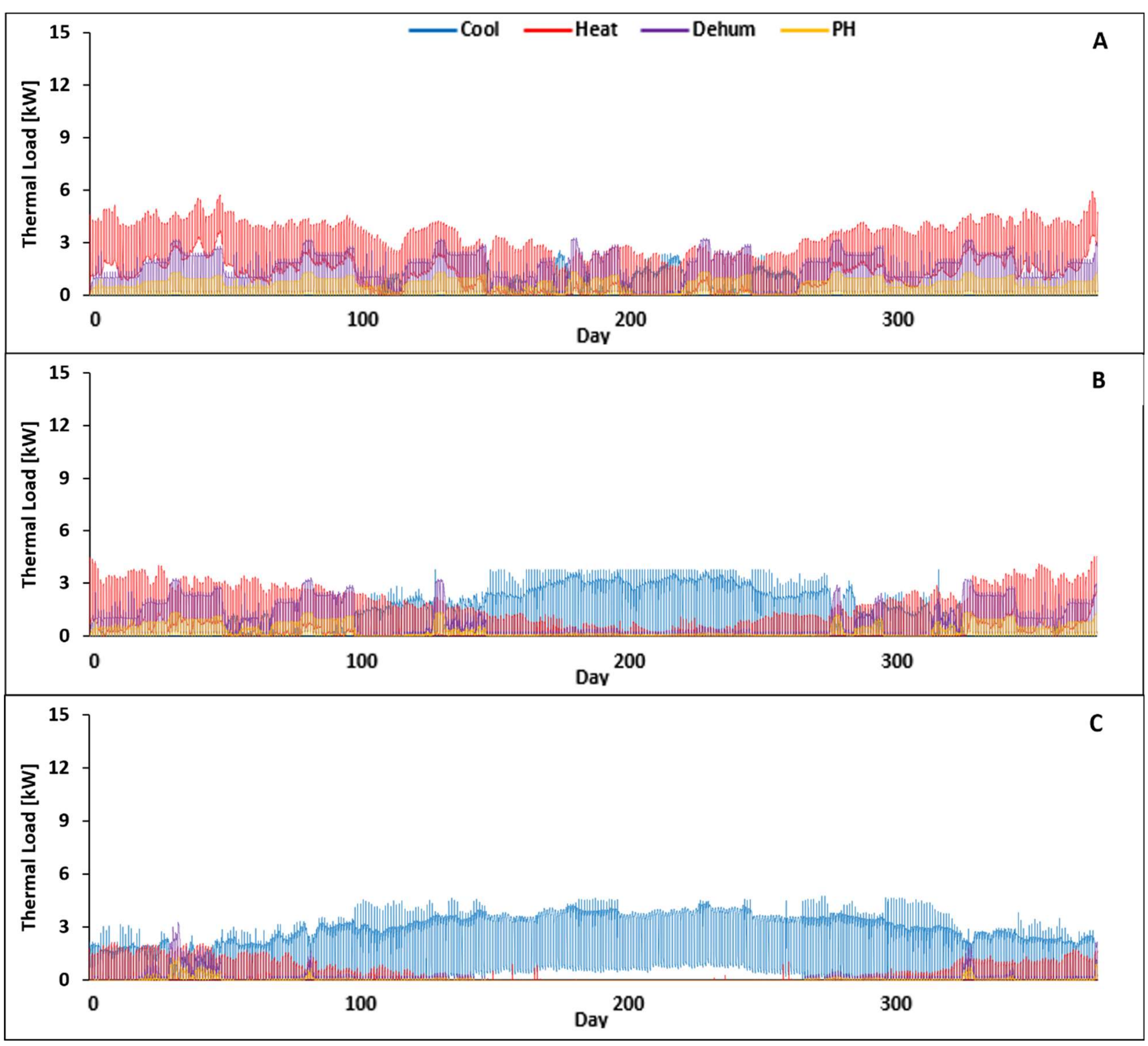

*Figure A11: Heating, cooling, dehumidification and post-heating loads throughout the year for a not-insulated VF in three locations: (A) Trondheim, (B) Shanghai, (C) Dubai. The results were obtained under conditions of 28 °C and 100 µmol m$^{-2}$ s$^{-1}$.*

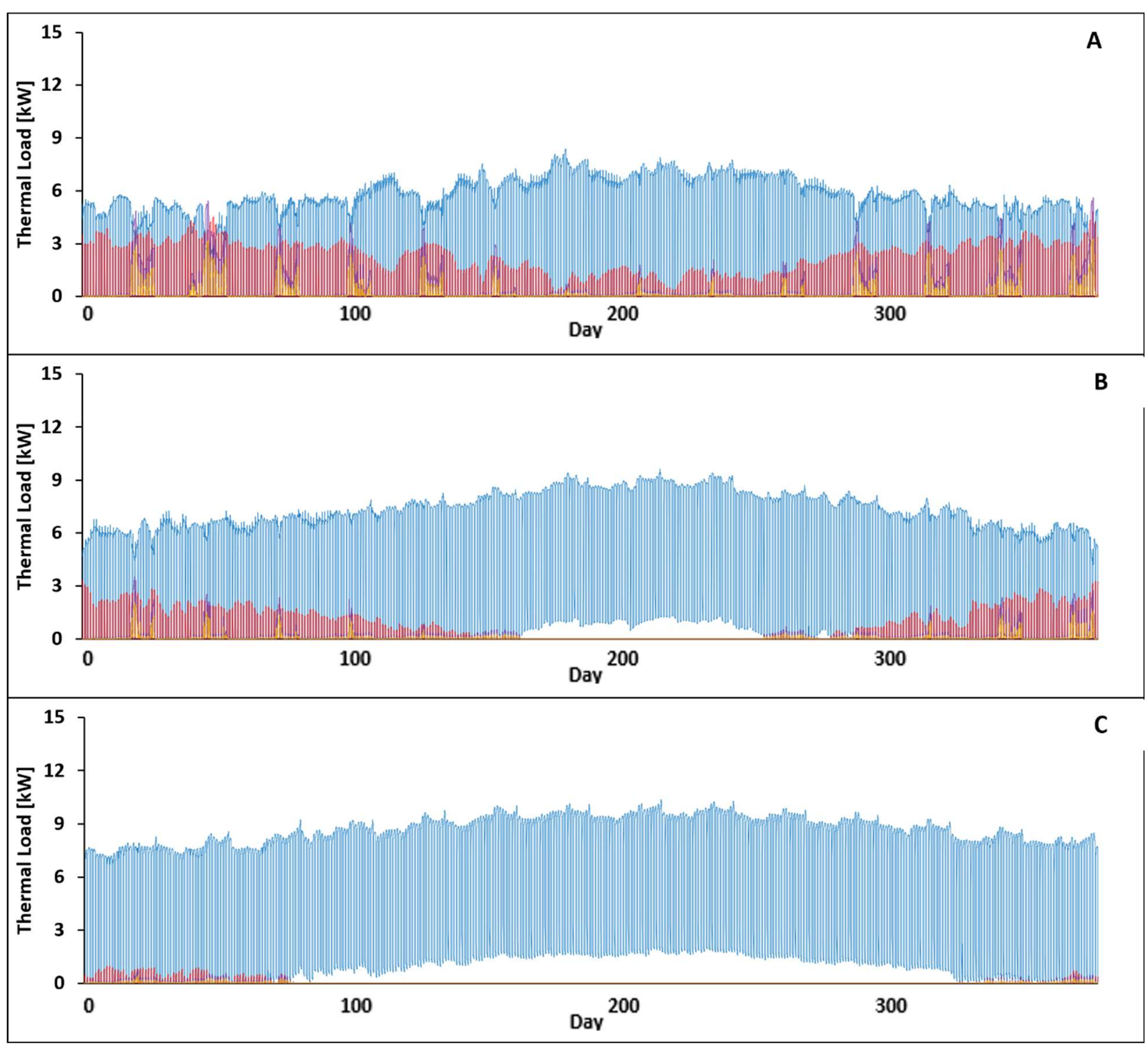

*Figure A12: Heating, cooling, dehumidification and post-heating loads throughout the year for a not-insulated VF in three locations: (A) Trondheim, (B) Shanghai, (C) Dubai. The results were obtained under conditions of 20 °C and 250 µmol $m^{-2}$ $s^{-1}$.*

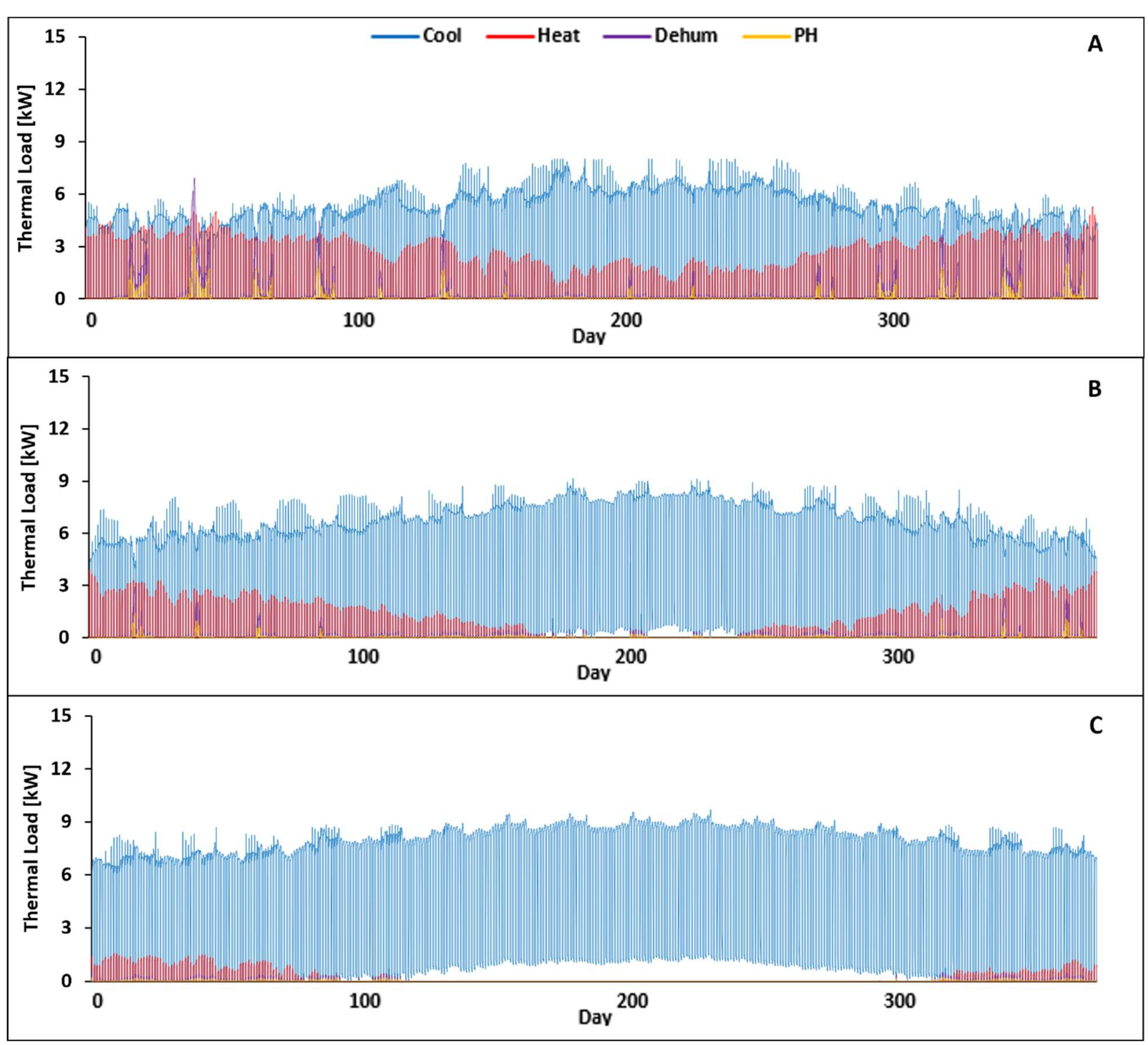

*Figure A13: Heating, cooling, dehumidification and post-heating loads throughout the year for a not-insulated VF in three locations: (A) Trondheim, (B) Shanghai, (C) Dubai. The results were obtained under conditions of 24 °C and 250 μmol m$^{-2}$ s$^{-1}$.*

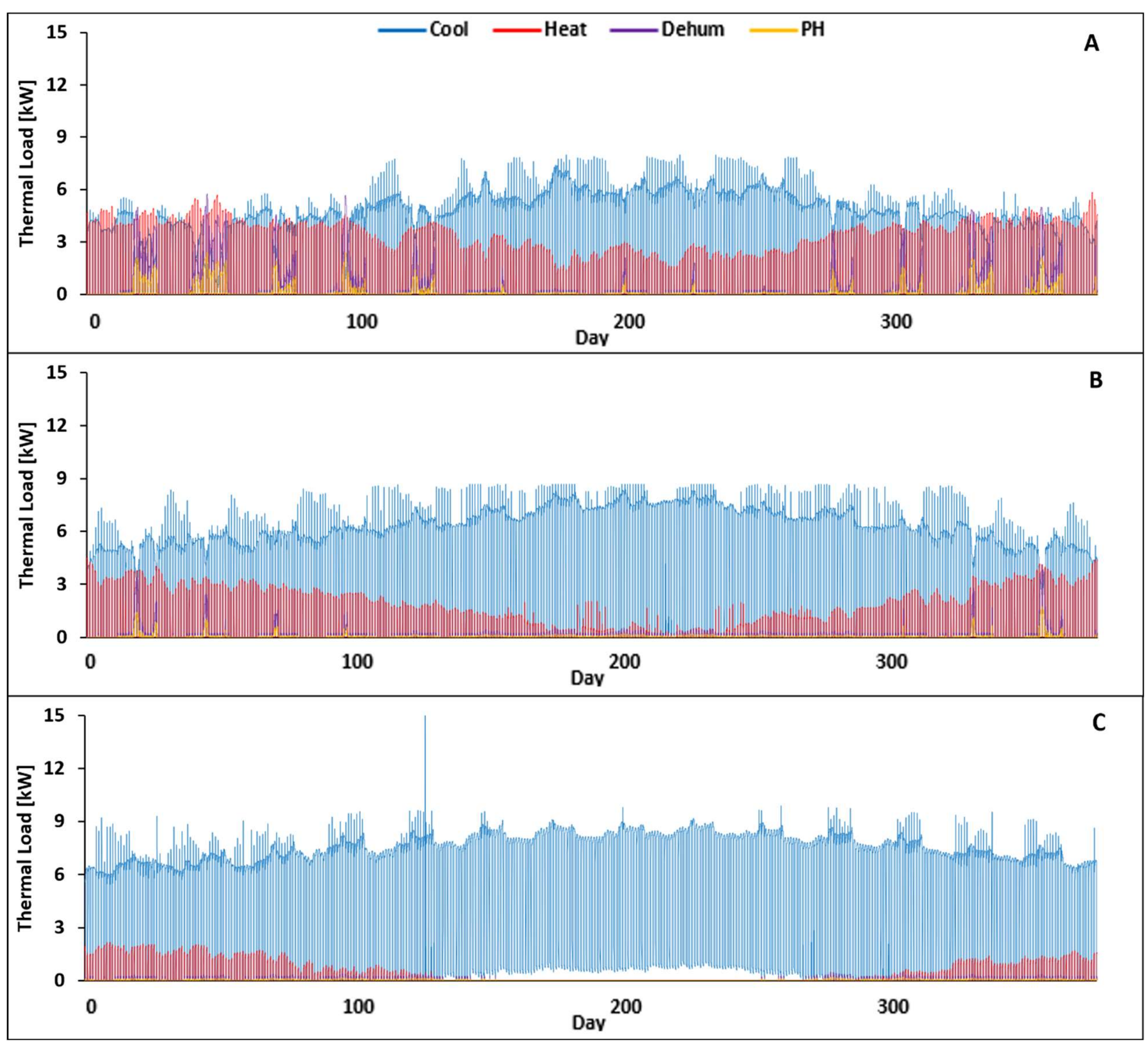

*Figure A14: Heating, cooling, dehumidification and post-heating loads throughout the year for a not-insulated VF in three locations: (A) Trondheim, (B) Shanghai, (C) Dubai. The results were obtained under conditions of 28 °C and 250 µmol $m^{-2}$ $s^{-1}$.*

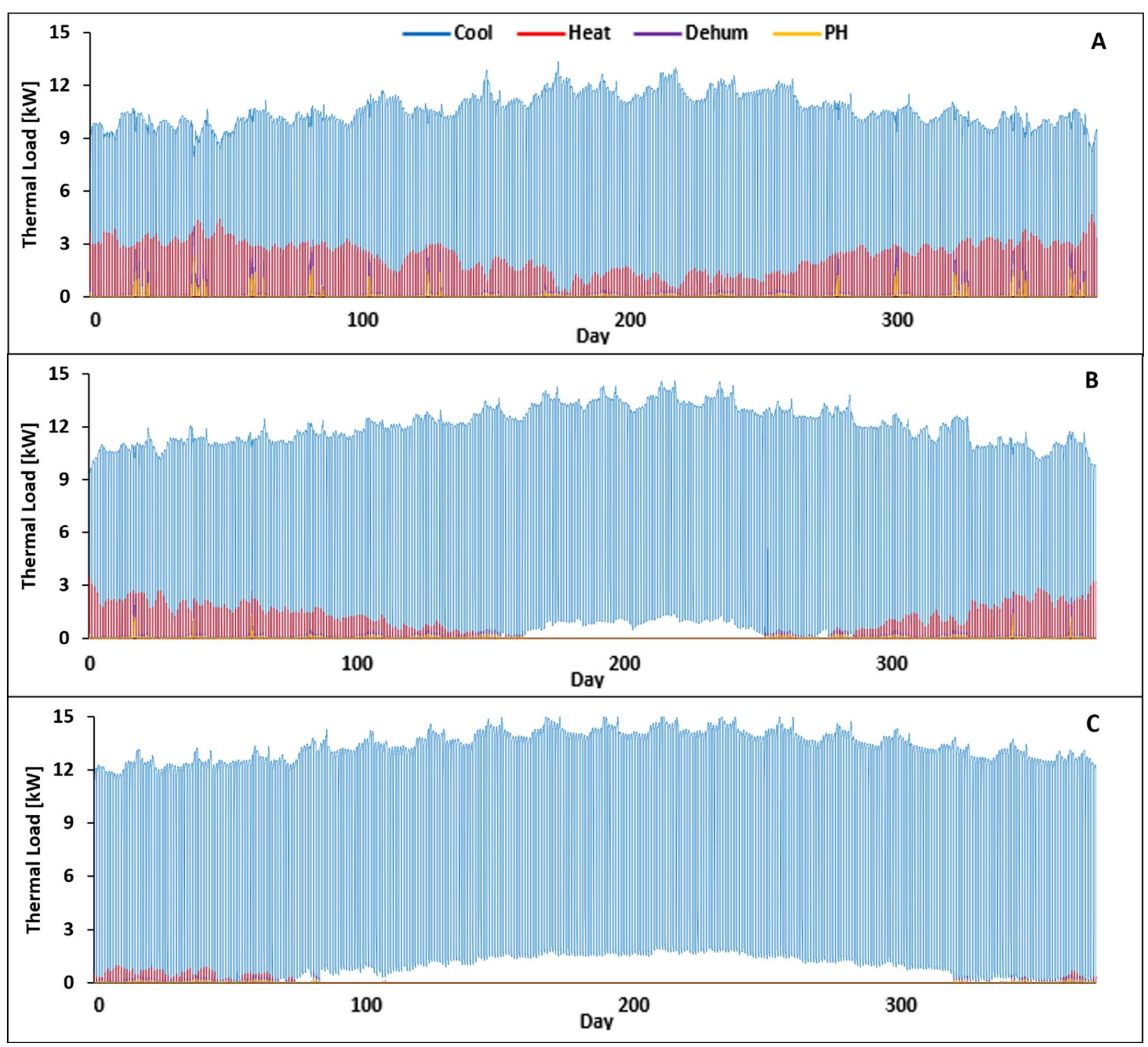

*Figure A15: Heating, cooling, dehumidification and post-heating loads throughout the year for a not-insulated VF in three locations: (A) Trondheim, (B) Shanghai, (C) Dubai. The results were obtained under conditions of 20 °C and 400 µmol m$^{-2}$ s$^{-1}$.*

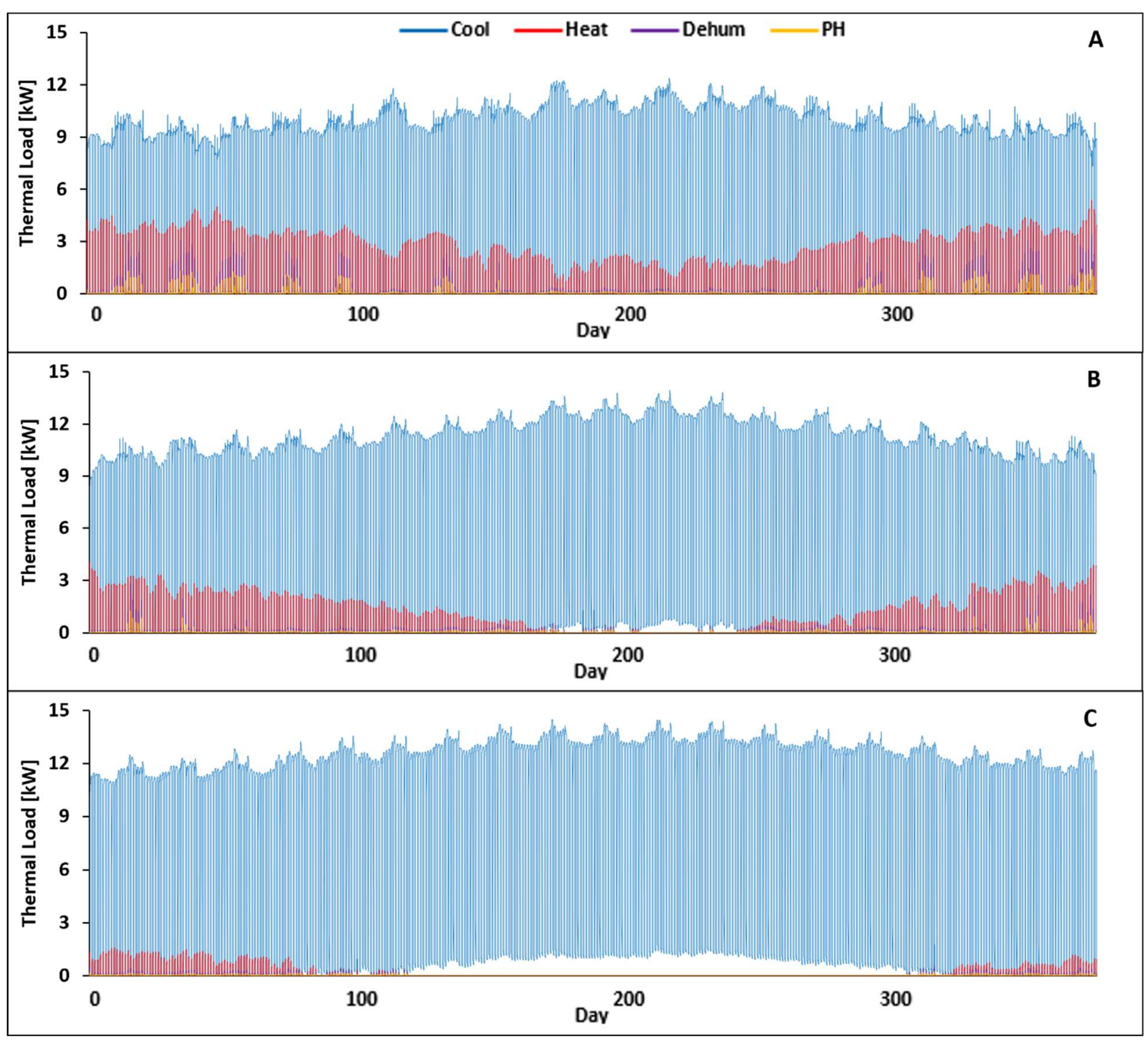

*Figure A16: Heating, cooling, dehumidification and post-heating loads throughout the year for a not-insulated VF in three locations: (A) Trondheim, (B) Shanghai, (C) Dubai. The results were obtained under conditions of 24 °C and 400 µmol $m^{-2}$ $s^{-1}$.*

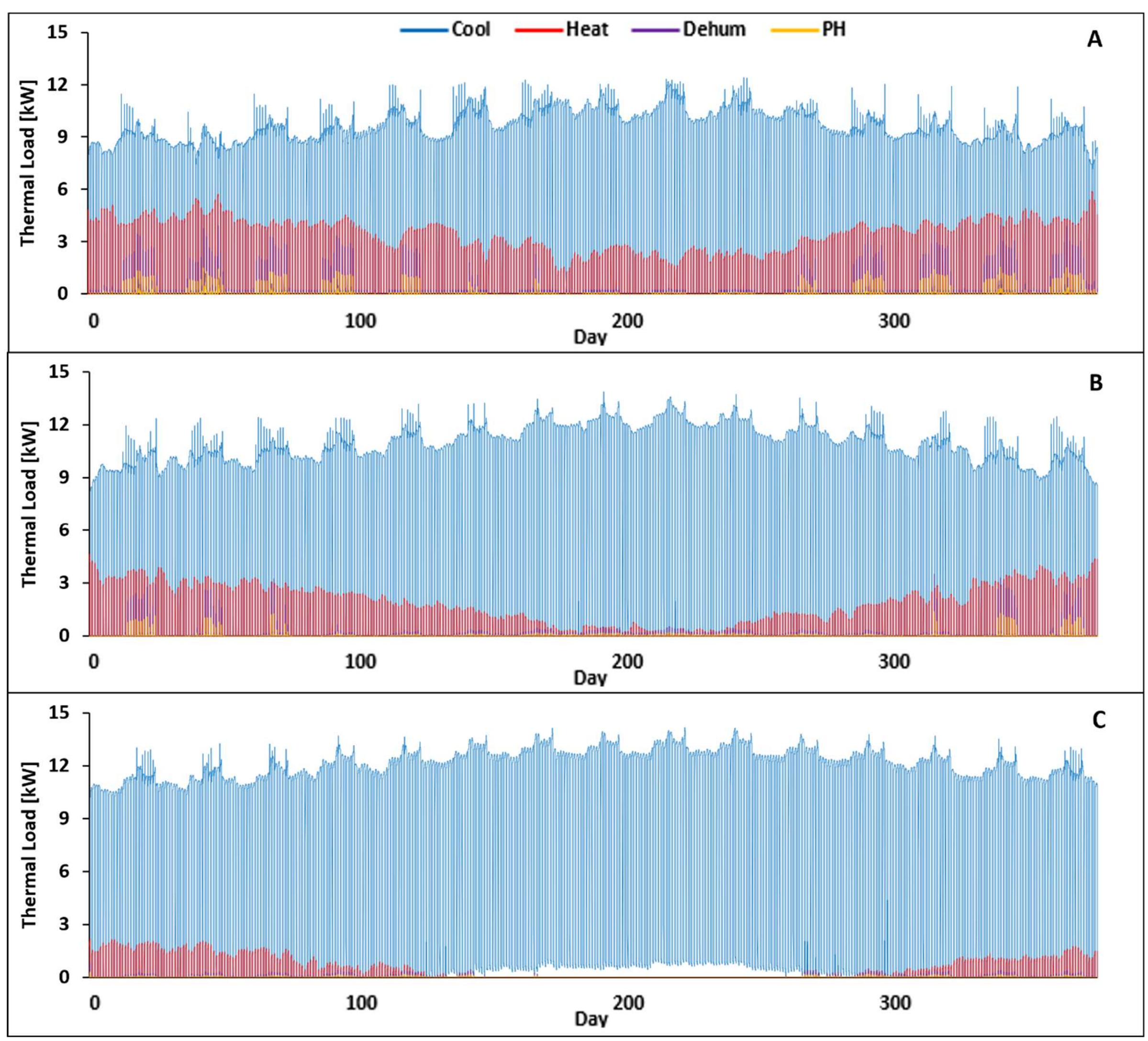

*Figure A17: Heating, cooling, dehumidification and post-heating loads throughout the year for a not-insulated VF in three locations: (A) Trondheim, (B) Shanghai, (C) Dubai. The results were obtained under conditions of 28 °C and 400 µmol $m^{-2}$ $s^{-1}$.*

**5.4 Specific Energy Consumption of Not-Insulated Vertical Farm**

This section presents the specific energy consumption of the non-insulated scenarios. The same assessments discussed in the main body of the paper remain valid for these cases as well. Notably, the absence of an insulation layer leads to an increase in heating demand, particularly in cold climates, which is partially offset by a reduction in cooling load. However, this reduction in cooling increases the VF's dehumidification and post-heating requirements, ultimately resulting in a higher specific energy demand, especially when the PPFD is low.

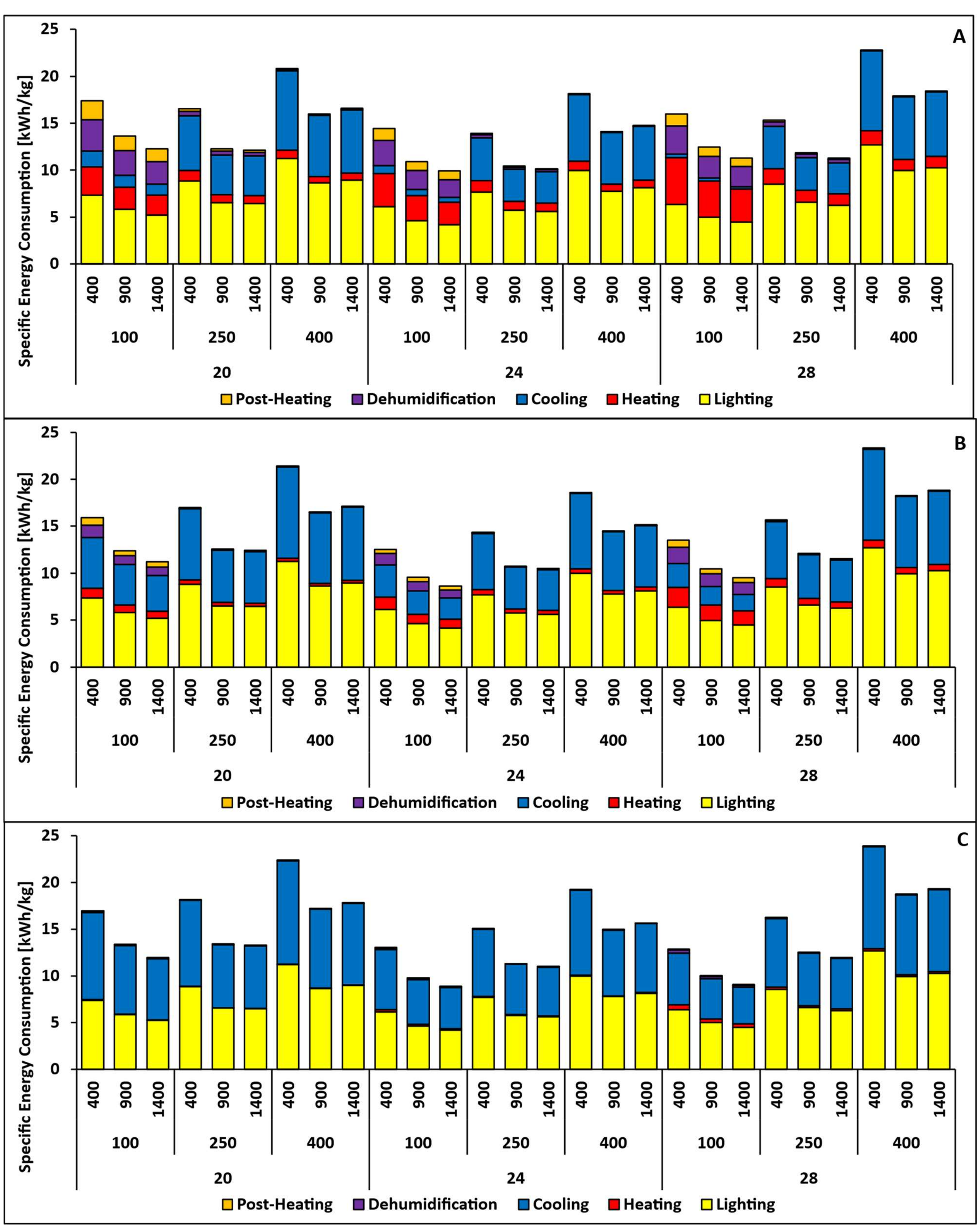

*Figure A18: Specific Energy Consumption of the not insulated VF under the three simulated locations: (A) Trondheim, (B) Shanghai, (C) Dubai.*

## 6. Declaration of generative AI and AI-assisted technologies in the writing process

During the preparation of this work, the authors used ChatGPT 4 in order to improve the readability of the manuscript. After using this tool/service, the authors reviewed and edited the content as needed and take full responsibility for the content of the published article.